\documentclass[aps,pra,10pt,twocolumn,longbibliography,nofootinbib,tbtags
]{revtex4-2}
\usepackage[utf8]{inputenc}
\usepackage{natbib}
\usepackage{graphicx}
\usepackage{xcolor}
\usepackage{mathtools}
\usepackage{amsmath}
\usepackage[colorlinks, linkcolor=blue]{hyperref}
\hypersetup{colorlinks,allcolors=black}
\usepackage{amssymb}
\usepackage{txfonts}
\usepackage{tabularx,graphicx}
\usepackage{epstopdf}
\usepackage{latexsym}
\usepackage{color, colortbl}
\usepackage{psfrag}
\usepackage{bbm}
\usepackage{bm}
\usepackage{titlesec}
\usepackage{dsfont}
\usepackage{feynmp}
\usepackage{slashed}
\usepackage{multirow}
\usepackage[normalem]{ulem}
\renewcommand{\vec}[1]{\boldsymbol{#1}}

\def \beq {\begin{eqnarray}}
\def \eeq {\end{eqnarray}}

\begin{document}
\title{Vacancy-assisted superfluid drag}
\author{Thomas G. Kiely}
\email{thomaskiely@ucsb.edu}
\affiliation{Kavli Institute for Theoretical Physics, University of California, Santa Barbara, California, USA.}
\author{Chao Zhang}
\email{chaozhang@ahnu.edu.cn}
\affiliation{
Department of Physics, Anhui Normal University, Wuhu, Anhui 241000, China
}
\author{Erich J. Mueller}
\email{em256@cornell.edu}
\affiliation{Department of Physics, Cornell University, Ithaca, New York, USA.}

\begin{abstract}
    We study superfluid drag in the two-component Bose-Hubbard model with infinitely strong repulsive interactions. { In this system, all transport is mediated by the motion of empty sites, or ``holes", and it is hard to move one component without moving the other.} We demonstrate, with a combination of analytic and numeric techniques, that the motion of holes leads to strong dissipationless coupling between currents in the two components.
    This behavior is attributable to polaronic correlations that emerge in the presence of spin currents, which can be observed in experiments.  
    { We derive a closed-form expression for the coupling on various lattices in arbitrary spatial dimensions, which we verify through numerical simulations on two dimensional lattices.}
\end{abstract}

\maketitle

\section{Introduction}










In an interacting multi-component system, internal forces tend to oppose the relative motion of the constituents.
In classical fluids this is 
a dissipative ``drag" force. 
{ An analogous effect is seen in {multi-component} superfluids,
in the form of a dissipationless}
``superfluid drag" first recognized in the context of $^3$He and $^4$He mixtures~\cite{andreevbashkin}: { 
Inducing a superflow in the $^4$He also leads to one in the $^3$He. { This is sometimes termed the Andreev-Bashkin effect.} The magnitude of the coupling, however, is quite small {
both in helium and in dilute gases \cite{PhysRevA.72.013616,Fil2004,Nespolo_2017,karle2019,romito2021}}. The situation is more dramatic
for two-component bosons on a lattice \cite{kaurov2005,sellin2018,kuklov2004,kuklov2004b}.} 
Near the jamming limit of one particle per site  the superfluid drag becomes very strong. Using analytic and two different numerical methods, we determine the value of the drag coefficient for a hard-core two-component gas of lattice bosons.  We also explore the  resulting spin correlations.


Two-component lattice bosons display a rich set of superfluid phenomena~\cite{kuklov2004,kuklov2004b,isacsson2005,hu2009}. In the weakly interacting limit, each component acts as an independent superfluid with a well defined phase, denoted $\phi_\uparrow$ { and} $\phi_\downarrow$.
Strong attractive interactions can lead to a paired-superfluid 
 {\cite{kuklov2004b}}, where 
{
the energy only depends on the sum of the phases}: $\Phi=\phi_\uparrow+\phi_\downarrow$. Similarly, 
repulsive interactions can cause binding between
particles of one species and holes of the other, resulting in a counter-superfluid 
{
where the energy only depends on the phase difference $\varphi=\phi_\uparrow-\phi_\downarrow$} \cite{kuklov2003}. When the interactions are strong, these superfluid states compete with magnetically-ordered insulating states that arise at commensurate filling fractions~\cite{kuklov2004b,Altman_2003,hubener2009}. {
Near the ordering transitions drag effects become large.  }

{
We focus on the hard-core limit, where only a single particle can sit on a site.  At unit filling (sometimes referred to as `half-filling' in this context) the system is jammed, and the ground state is highly degenerate.  Adding a single hole breaks this degeneracy, and allows the  atoms to rearrange themselves, much like a children's puzzle where a missing tile allows one to rearrange a scrambled image.  As with the fermionic case, the energy is minimized by taking a symmetric combination of all possible patterns \cite{nagaoka1966,eisenberg2002}.  This is a $xy$ ferromagnet if one treats the two components as the $z$ component of a pseudospin.  Adding supercurrents frustrates this ferromagnetism, leading to a spin-polaron where the phases twist more rapidly near the location of the hole.}

{
This physics is readily studied by cold atoms in optical lattices.  The most natural realization involves using two different hyperfine states of a single atomic species, such as $^{87}$Rb.  Superfluid drag can be probed by the response to spin-dependent forces, by studying collective modes \cite{parisi,Nespolo_2017}, or the behavior of vortices \cite{dahl,karle2019,sellin2018}.  In cold gas microscopes one can even observe the 
{positions of individual atoms}, allowing one to directly measure how the spins twist near a hole \cite{Parsons2016,Cheuk2016,Boll2016,Koepsell2019,Koepsell2021,bohrdt}.
}


{
Calculating superfluid drag is challenging.}
At the level of mean-field theory, the drag coefficient is zero. Corrections due to quantum fluctuations can be treated using the Bogoliubov formalism~\cite{linder2009,hofer2012,hartman2018}
{ or the quantum Gutzwiller ansatz~\cite{colussi2022}, although such calculations can not reliably capture the behavior of strongly-correlated superfluids.}
In the strongly-interacting limit, the superfluid drag {
has been} 
computed by studying the 
{
statistics}
of world-lines in quantum Monte Carlo simulations~\cite{kuklov2004,kuklov2004b,kaurov2005,sellin2018} {
though the low hole density limit presented challenges}. 
More recently, Refs.~\cite{contessi2021,gremaud2021} studied superfluid drag in one-dimensional models using tensor network methods.

We build off {
of all} these {
prior} works, studying superfluid drag in the limit of hard-core interaction on two-dimensional lattices via {
three complementary techniques.  We present an analytic calculation that allows us to calculate the exact drag coefficient in the limit of vanishing hole density { (in any dimension)}.  We supplement this with infinite matrix product state calculations on cylinder geometries with transverse size $C=3,4,5$.  These allow us to consider finite hole density, but require extrapolating to $C\to\infty$.  Finally, we use a quantum Monte-Carlo approach to calculate the drag in finite torus geometries of  $L\times L$ sites with $L\sim 20$.  

All three techniques have different regimes where they are most reliable.  The analytic calculation only works at vanishing hole density.  Conversely, the Monte Carlo calculations are most reliable at large hole densities.  The tensor network calculations bridge these two limits, but have significant finite size effects.  With appropriate extrapolation techniques, we find that all three approaches agree with one-another.  Thus we are able to present a well-validated global view.}  



\section{Superfluid drag}
The low-energy properties of a multi-component superfluid are dominated by phase fluctuations in the order parameters for each species~\cite{kaurov2005}. For a single-component superfluid, the energetic cost of phase gradients is set by the superfluid stiffness.   
Supercurrents are proportional to the phase gradient, and one can interpret 
the superfluid stiffness 
as the density of 
particles that participate in dissipationless flow \cite{kiely2022}.
For a multi-component system, there is a matrix of stiffnesses, e.g. {
the energy can be written}
\begin{equation}\label{estiff}
    E=E_0+\frac{1}{2}{ \frac{\hbar^2}{m}}\int dx~\begin{pmatrix} \nabla \theta_\uparrow & \nabla \theta_\downarrow \end{pmatrix}\begin{pmatrix} \rho_{\uparrow\uparrow} & \rho_{\uparrow\downarrow} \\ \rho_{\uparrow\downarrow} & \rho_{\downarrow \downarrow}\end{pmatrix}\begin{pmatrix}\nabla \theta_\uparrow \\ \nabla\theta_\downarrow \end{pmatrix},
\end{equation}
{
where $E_0$ is a constant which does not depend on the phase gradients.}
If the different spin species interact with one another, the off-diagonal elements of the stiffness matrix will generically be finite. The existence of finite off-diagonal elements {
means that the current in one component depends on the phase twists of both components, $j_\uparrow\propto \partial E/\partial \nabla \theta_\uparrow$ $\propto\rho_{\uparrow\uparrow} \nabla \theta_{\uparrow}+\rho_{\uparrow\downarrow}\nabla \theta_{\downarrow}$.  Applying a force to one component would twist its phase, generating currents in both.
This phenomenon}
is what we refer to as {\it superfluid drag}.

In the two-component system, {
the off-diagonal} coefficient, $\rho_{\uparrow\downarrow}$, controls the magnitude of the current response in one component due to a phase twist in the other component. For this reason, $\rho_{\uparrow\downarrow}$ also controls the magnitude and sign of the interaction between vortex excitations of different species~\cite{sellin2018}.
The drag coefficient, ${
\kappa=\rho_{\uparrow\downarrow}/
\sqrt{\rho_{\uparrow\uparrow} \rho_{\downarrow\downarrow}}}$, is a dimensionless 
{
number} that 
{
quantifies} the relative importance of the superfluid drag in a given system.
{
We are considering the symmetric case where $\rho_{\uparrow\uparrow}=\rho_{\downarrow\downarrow}$.}

\section{Model}
We consider the two-component Bose-Hubbard model on a two-dimensional (2D) square lattice:
\begin{equation}\label{eq:ham_bh}
    H=-t\sum_{\langle i,j\rangle,\sigma}\left(a^\dagger_{i\sigma}a_{j\sigma}+{\rm H.c.}\right)+\frac{U}{2}\sum_{i,\sigma,\tau}n_{i\sigma}n_{i\tau}-\mu\sum_{i\sigma}n_{i\sigma},
\end{equation}
where $a_{i\sigma},a_{i\tau}$ are the bosonic annihilation for spin species $\sigma,\tau=\uparrow,\downarrow$ on site $i$, 
$n_{i\sigma}=a^\dagger_{i\sigma}a_{i\sigma}${
, and $\mu$ is the chemical potential}. Note that Eq.~(\ref{eq:ham_bh}) has an ${\rm SU}(2)$ spin symmetry: the Hamiltonian is invariant under rotations $U_s$ that mix the spin species. One might also consider models that break this symmetry, e.g. if the inter-species interaction strength differs from the on-site interaction strength. 
This distinction is irrelevant in the strongly-interacting limit considered here.

In the limit $t/U\to 0$, the
interaction term can be replaced by a hard-core constraint, which can formally be treated as an operator identity,
$ a_{i\sigma}a_{i\tau} =0$. For finite but small $t/U$, the leading-order correction to this hard-core model is a nearest-neighbor ferromagnetic Heisenberg interaction with coefficient $J=4t^2/U$.
At unit-filling, this {
typically} implies that the ground state is 
{ Mott insulator with XY ferromagnetic order}
, though
other magnetic orders can be found at unit filling if one breaks the ${\rm SU}(2)$ spin symmetry~\cite{Altman_2003}.
{
We will not include such superexchange terms in our modeling, restricting ourselves to the $t/U\to 0$ limit.   At unit filling the ground state will be highly degenerate, but adding holes will break the degeneracy, leading to a ferromagnetic ground state.}
{
This phenomenon}
is known as {\it kinetic magnetism}:
the kinetic energy of itinerant holes is maximized {
when the spin wavefunction is fully symmetric}~\cite{nagaoka1966}. { Crucially, for $t/U\ll1$, this implies that the doped ground state will always be ferromagnetic, irrespective of the magnetic order at unit filling.}
{
We parameterize the particle number by the density of vacancies, $x_v=1- n_\uparrow-n_\downarrow$.} 

 In this lattice model, it is natural to make the substitution $\hbar^2/m\to 2t{ a^2}$ in Eq.~(\ref{estiff}) { where $a$ is the lattice spacing.  We also work in units where $a=1$.} { In doing so, 
 the superfluid densities 
 become dimensionless numbers.}

\section{Variational Formulation}\label{var}
{
Our primary strategy  involves using a variational principle to calculate the superfluid density tensor.  The energy in Eq.~(\ref{estiff}) can be found by minimizing the energy in Eq.~(\ref{eq:ham_bh}) with the constraint that $\theta_\uparrow$ and $\theta_\downarrow$ have a fixed profile.  In particular, we consider a linear phase gradient where $\nabla \theta_\uparrow={\vec k_\uparrow}$ and $\nabla \theta_\downarrow={\vec k_\downarrow}$ are constant in space.  To simplify the arithmetic, we introduce a gauge transformation which effectively places us in the frame of the moving superfluid.  In particular we define 
$\tilde H=UH U^\dagger$, with
\begin{equation}\label{rot}
    U(\vec k,\vec q)=\prod_{j}\exp\left[i\vec r_j\cdot\left(\vec k(n_{j\uparrow}+n_{j\downarrow})+\vec q(n_{j\uparrow}-n_{j\downarrow})\right)\right]
\end{equation}
corresponding to a spin-dependent Galilean boost.
Here ${\vec k}=(\vec k_\uparrow+\vec k_\downarrow)/2$  relates to uniform currents where both components flow in the same direction.  Conversely,  ${\vec q}=(\vec k_\uparrow-\vec k_\downarrow)/2$ corresponds to  counterflow.  The vector $\vec{r}_j$ gives the $x$ and $y$ coordinates of the $j$'th site, which are chosen to sit on a square grid with unit spacing.

The transformed Hamiltonian (in the hard-core limit) is
\begin{equation}\label{twistH}
\tilde H = -\sum_{\langle i,j\rangle,\sigma} t_{ij,\sigma}\left(a_{i\sigma}^\dagger a_{j\sigma} +{\rm H.c.}\right)-\mu\sum_{i\sigma} n_{i\sigma},
\end{equation}
with $t_{ij,\uparrow}=t e^{i{(\vec k+\vec q)\cdot (\vec r_i-\vec r_j)}}$ and $t_{ij,\downarrow}=t e^{i{(\vec k-\vec q)\cdot (\vec r_i-\vec r_j)}}$.  We then must find 
$E(k,q)={\rm min}_{\{|\psi\rangle\}}\langle\psi|\tilde H|\psi\rangle$ where the minimization is constrained to be over the set of translationally-invariant wavefunctions $\{|\psi\rangle\}$. 

As $k,q\to0$, the energy of the resulting current-carrying state is given by
\begin{equation}\label{eq:twistenergy}
    (E(k,q)-E_0)/t=\rho_+k^2+\rho_-q^2+\ldots
\end{equation}
where $E_0$ is the ground-state energy and $t$ is the tunneling matrix element. The coefficients $\rho_\pm=(\rho_{\uparrow\uparrow}+\rho_{\downarrow\downarrow}\pm2\rho_{\uparrow\downarrow})/2$ are the superfluid densities in the density ($+$) and spin ($-$) channels. These are the eigenvalues of the superfluid density matrix, and 
the drag coefficient is $\kappa=(\rho_+-\rho_-)/(\rho_++\rho_-)$.

We again emphasize that the energy minimization must be taken with the constraint that $|\psi\rangle$ is transitionally invariant.  Since $\tilde  H$ is gauge equivalent to $H$, the unconstrained minimum would result in $E=E_0$, independent of $k$ and $q$. 
}

{
We carry out the constrained minimization in two ways.  In Sec.~\ref{sec:pert} we find the exact minimum in the limit of a single hole.  In Sec.~\ref{MPS} we use the Variational Uniform Matrix Product State (VUMPS) method to 
numerically optimize a translationally invariant matrix product state.  
Conversely, the Monte-Carlo calculations in Sec.~\ref{mc}  do not rely upon this variational formulation, but instead extract the superfluid density from the statistics of world line winding numbers.
}

\section{Perturbation theory in the single-hole limit\label{sec:pert}}
As noted earlier, the ground state of Eq.~(\ref{eq:ham_bh}) at unit-filling, zero magnetization and 
$U=\infty$ is
{
highly degenerate.  This degeneracy is broken by adding a hole, in which case the ground state becomes}
an easy-plane (XY) ferromagnet. 
{
We} pick a preferred in-plane direction and write the 
{
unit filled parent} state as $|\phi\rangle=\prod_j b^\dagger_{j+}|0\rangle$ where $|0\rangle$ is the vacuum and $b^\dagger_{j\pm}=(a_{j\uparrow}^\dagger\pm a_{j\downarrow}^\dagger)/\sqrt{2}$ 
{
creates a boson in a superposition of the two spin states. } 
{
This is an eigenstate of $S_{xj}= (a_{j\uparrow}^\dagger a_{j\downarrow}+a_{j\downarrow}^\dagger a_{j\uparrow})/2$, which is the $x$ component of the local spin operator $\vec S_j$. 
The ground state with a single hole is  $|\psi_0\rangle= (1/\sqrt{N_s})\sum_j b_{j+}|\phi\rangle$, where $N_s$ is the number of sites.
One can readily verify that $|\psi_0\rangle$ is an eigenstate of 
Eq.~(\ref{eq:ham_bh})
}. 

{
We now add phase twists, as detailed in Sec.~\ref{var}.}  The Peierls phases
break the ${\rm SU}(2)$ invariance of the kinetic term 
and in  our rotated basis the Hamiltonian $\tilde H$ 
takes the form
\begin{equation}\label{eq:rotatedH}
    \begin{split}
        \tilde H/t=&-\sum_{\langle i,j\rangle}\left(b^\dagger_{i+}b_{j+}+b^\dagger_{i-}b_{j-}\right)e^{i\vec r_{ij}\cdot \vec k}\cos\left(\vec r_{ij}\cdot \vec q\right)+{\rm H.c.}\\
    &-i\sum_{\langle i,j\rangle}\left(b^\dagger_{i+}b_{j-}+b^\dagger_{i-}b_{j+}\right)e^{i\vec r_{ij}\cdot \vec k}\sin\left(\vec r_{ij}\cdot \vec q\right)+{\rm H.c.}
    \end{split}
\end{equation}
where $\vec r_{ij}=\vec r_i-\vec r_j$. {
We have omitted the chemical potential as we will explicitly conserve particle number, making it unneccessary.  As before, the hard-core constraint} 
implies $b^\dagger_{i\sigma}b^\dagger_{i\tau}=0$, with $\sigma,\tau=\pm${, so the terms in Eq.~(\ref{eq:rotatedH}) only allow a particle to hop onto another site if that site is originally empty.  The second line corresponds to hopping events in which the particle's spin flips.} 
In what follows, we will perform perturbation theory in $\vec k$ and $\vec q$ in order to determine the energy to $\mathcal{O}(k^2,q^2)$.

{
As already argued, } when $\vec q=0$ {
and $\vec k=0$,} the single-hole ground state of Eq.~(\ref{eq:rotatedH}) is 
$|\psi_0\rangle=(1/\sqrt{N_s})\sum_jb_{j+}|\phi\rangle$. 
{
This function is an eigenstate of the first term in Eq.~(\ref{eq:rotatedH}), and it remains the lowest energy uniform state when $\vec k\neq 0$ and $\vec{q}=0$.}
The resulting superfluid density 
is $\rho_+=\bar n_+(1-\bar n_+)$.  This result coincides with what one expects from mean-field theory.

{
When $\vec q \neq 0$, $\tilde H$ contains terms where a $+$ spin sitting beside a hole can be converted into a $-$ spin { while hopping onto the empty site}.  To leading order in $\vec q$ we only need to include a single such flipped spin, and can make the ansatz that the uniform ground state wavefunction takes the form}
\begin{equation}\label{eq:flipansatz}
    |\psi\rangle=\frac{1}{\sqrt{N_s}}\sum_jb_{j+}\left(f_0+\sum_{s\neq 0}f_sb_{j+s,-}^\dagger b_{j+s,+}\right)|\phi\rangle
\end{equation}
where $f_{s\neq 0}\propto q$ {
is small.  One recognizes $f_0$ as the amplitude to have no flipped spins.  The coefficient  $f_s$ is the amplitude to have a single   flipped spin, separated from the hole by the 
vector $s$. 
Contributions to the wavefunction involving more flipped spins will be suppressed by higher powers of $q$.} 

Given {
the wavefunction in Eq.~(\ref{eq:flipansatz}),} 
we now solve the Schr\"odinger equation $\tilde H|\psi\rangle=E|\psi\rangle$ to determine $E(k,q)$. To simplify notation, we introduce
\begin{eqnarray}
t_{
{
\vec{u}}} &=& te^{i {\vec k \cdot \vec{u}}}\cos {\vec q\cdot \vec{u}}\\
\lambda_{\vec u} &=& te^{i {\vec k \cdot \vec u}}\sin {\vec q\cdot \vec u}\\
\epsilon_{\vec p} &=& -\sum_{\vec u} t_{\vec u} e^{-i\vec p\cdot \vec u}\\
\eta_{\vec p} &=& -i\sum_{\vec u} \lambda_{\vec u} e^{-i\vec p\cdot \vec u}
\end{eqnarray}
where the sum over $\vec u$ denotes a sum over nearest-neighbors on the square lattice. The Schr\"odinger equation yields coupled equations for the coefficients $f_0$ and $f_{\vec s}$:
\begin{eqnarray}
E f_0 &=& \epsilon_0 f_0 -i \sum_{\vec u} \lambda_{\vec u} f_{-\vec u}\label{f0eqn}\\
E f_{\vec s} &=& \sum_{
\vec u}\left[
-t_{\vec  u} (1-\delta_{\vec s,\vec u}) f_{\vec s-\vec u}-\delta_{\vec s,\vec u} (t_{\vec u} f_{-\vec u} +i\lambda_{\vec u} f_0
)
\right]\label{fseqn}
\end{eqnarray}
Our interpretation of Eq.~(\ref{fseqn}) is that the flipped spin (initially displaced by $\vec s$ from the hole in Eq.~(\ref{eq:flipansatz})) can hop around like a normal particle, except that it can hop {\it over} the hole, i.e. hopping from $\vec u$ to $-\vec u$. This correlated hopping process is the origin of the coupling between density and spin degrees of freedom in the single-hole calculation. {
There is also a process where the flipped spin can be created/destroyed if it is on a site neighboring the hole.  The matrix element for this process is $-i\lambda_{\vec u}$.}

We solve Eqns.~(\ref{f0eqn}) and (\ref{fseqn}) by making a Fourier expansion,
\begin{eqnarray}
f_{\vec s\neq 0} &=& \frac{1}{\sqrt{N_s}} \sum_{\vec p} e^{i \vec p\cdot \vec s} g_{\vec p} \\
g_{\vec p} &=&\frac{1}{\sqrt{N_s}} \sum_{\vec s \neq 0} e^{-i\vec p\cdot \vec s} f_{\vec s} + c,
\end{eqnarray}
where the constant $c$ {
may be chosen arbitrarily.  This freedom reflects the fact that this Fourier expansion contains one more degree of freedom than the coefficients $f_{\vec s}$ with $s\neq 0$. } 
Summing over $\vec s$ in Eq.~(\ref{fseqn}), we find
\begin{equation}
    (E-\epsilon_{\vec p})(g_{\vec p}-c) =
\frac{1}{\sqrt{N_s}}\sum_{\vec u} \left[1-e^{-i \vec p\cdot \vec u}\right] t_{\vec u} f_{-\vec u}
+\frac{1}{\sqrt{N_s}}\eta_{\vec p} f_0.
\end{equation}
{Notably, the right-hand side depends only on $f_0$ and $f_{ \vec u}$ for $\vec u$ a nearest-neighbor of zero.} We can then write an arbitrary $f_{\vec s}$ in terms of these via
\begin{eqnarray}
f_{\vec s} &=& \frac{1}{\sqrt{N_s}}\sum_{\vec p} e^{i \vec p\cdot \vec s} (g_{\vec p}-c)\\
&=&
\sum_{\vec u} t_{\vec u} f_{-\vec u} (\Lambda_{\vec s} -\Lambda_{\vec s-\vec u})
+ f_0 \Gamma_{\vec s}
\label{fs}
\end{eqnarray}
where we have defined
\begin{eqnarray}
\Lambda_{\vec s} &=& 
\frac{1}{N_s} \sum_{\vec p} \frac{e^{i\vec p\cdot \vec s}}{E-\epsilon_{\vec p}}\label{eq:lambda}\\
\Gamma_{\vec s} &=& \frac{1}{N_s} \sum_{\vec p} \frac{e^{i \vec p\cdot \vec s} \eta_{\vec p}}{E-\epsilon_{\vec p}}\label{eq:gamma}
\end{eqnarray}
Given this form of $f_{\vec s}$, we can fix all the free parameters in Eq.~(\ref{eq:flipansatz}) by solving Eq.~(\ref{f0eqn}) and Eq.~(\ref{fseqn}) for all ${\vec s}$ which are a neighbor of zero. On a regular lattice with coordination number $z$, this yields $z+1$ coupled equations. {
These are solved in Appendix~\ref{sec:pert_app} resulting in}
%
\begin{equation}\label{eq:pertenergy}
    E(k,q)/t=-4+k^2+\left(\frac{1-\kappa}{1+\kappa}\right)q^2+\ldots
\end{equation}
where $\kappa$, the superfluid drag coefficient, is given by 
\begin{eqnarray}\label{kappaint}
    \kappa
    &=&\int \frac{d p_x\,d p_y}{(2\pi)^2}\frac{\cos 2p_x-1}{-4+2\cos p_x +2 \cos p_y}\\
    &=&1-2/\pi.
\end{eqnarray}
We emphasize that Eq.~(\ref{eq:pertenergy}) should be understood as the energy density {\it per hole} when comparing to a result at finite hole density $x_\nu=1-\sum_\sigma \bar n_\sigma$. Comparing with Eq.~(\ref{eq:twistenergy}), we find
\begin{align}
    \lim_{x_\nu\to 0}~\rho_+&=x_\nu &
    \lim_{x_\nu\to 0}~\rho_-&=\frac{x_\nu}{\pi-1}.
\end{align}
As 
{
already discussed}, $\rho_+$ 
is consistent with the {
low $x_v$ limit of the} mean-field result $\rho_{+,{\rm MF}}=x_\nu(1-x_\nu)$. 
At the level of mean-field theory, one would expect $\rho_-=\rho_+$, i.e. a drag coefficient $\kappa=0$. 
The depletion of the superfluid density in the spin channel is the consequence of strong superfluid drag induced by the correlated hopping term between the flipped spin and the single hole.

We conclude this section by noting that the same calculation straightforwardly applies to single-hole limits of other regular lattices. In one dimension, we find a drag coefficient $\kappa_{1D}=1$, which is consistent with kinetic constraints: particles of different spin cannot move around one another, so $\rho_-=0$. On an $n$-dimensional hypercubic lattice, we find that $\kappa_{nD}\sim 1/(z-1)$ where $z=2n$ is the coordination number.
On the 2D triangular lattice, we find $\kappa_{\Delta}\approx0.23$. 
These results 
are consistent with the picture
that lattices with greater coordination number are more ``weakly-interacting." Indeed, as the origin of superfluid drag in this limit is the correlated-hopping term, increasing $z$ creates more opportunities for the spin to hop ``normally," thus decreasing the superfluid drag.

{
For comparison with our numerics we can also consider finite $L\times L$ lattices with periodic boundary conditions, or infinite cylinders of circumference $C$.  In the former, the integral over $k_x$ and $k_y$ in Eq.~(\ref{kappaint}) are replaced with sums where $k_x=2\pi n_x/L$, and $k_y=2\pi n_y/L$ where $n_x,n_y$ are integers between $0$ and $L-1$.  For infinite cylinders, the $k_x$ integral remains, but the $k_y$ integral is replaced by a sum. 
{
For $C=3,4,5$ we find $\kappa=2-\sqrt{7/3},~~2-(1/2)\sqrt{5+2\sqrt{6}},~~2-(1/5)\sqrt{35+2\sqrt{205}}$.}
}
{
We emphasize that these results are {\em exact}, and do not rely upon any approximations.}

\section{Matrix product states}\label{MPS}
{
The approach in Sec.~\ref{sec:pert} does not readily generalize to larger hole density, $x_v$.  Thus we turn to a numerical approach.  We make the ansatz that $|\psi\rangle$ can be written as a translationally invariant matrix product state \cite{Schollwck2011}.  We then minimize the energy over the space of all such states.

For these calculations we consider
an infinite cylinder geometry, taking cylinders of circumference $C=3,4,5$.  The sites of this lattice are then arranged in one dimensional line \cite{liang1994}.  Thus our ansatz is a uniform matrix product state with a unit cell of $C$ sites.  We use the VUMPS algorithm \cite{VUMPS} to minimize the energy at fixed bond dimension $\chi$, corresponding to the largest size of the matrices in the ansatz.  As $\chi$ is made larger the ansatz becomes more expressive.  We should recover the exact result in the limit $\chi\to\infty$.

We first perform the minimization with $\vec k =\vec q=0$.  
We then repeat the calculation for a series of small $\vec k$ and $\vec q$, which are oriented along the long axis of the cylinder{
, taken to be the x-axis}.}
For sufficiently small twists, we converge to the lowest-energy translationally-invariant ground state, from which we
extract $\rho_\pm$ via Eq.~(\ref{eq:twistenergy}). We previously used {
the same procedure} to determine the superfluid density of the 1D single-component Bose-Hubbard model~\cite{kiely2022}. {
As in that work, here it is essential to not constrain the total density or magnetization using conserved quantum numbers~\cite{kiely2022b} -- one may readily show that the energy of $\tilde H(\vec k,\vec q)$ with respect to a spin and density-conserving iMPS is independent of $\vec k$ and $\vec q$.}
{
{
We repeat our calculations for a sequence of $\chi$ and extrapolate to $\chi=\infty$, a procedure referred to as finite entanglement scaling \cite{kiely2022,pollmann2009}. }
For $C=3$ we used $\chi\leq120$, for $C=4$ we used $\chi\leq140$ and for $C=5$ we used $\chi\leq160$.}

\section{Monte Carlo}\label{mc}

The Matrix Product State calculations in Sec.~\ref{MPS} are limited to cylinders of modest size, so we supplement them with quantum Monte-Carlo calculations using the Worm algorithm~\cite{PROKOFEV1998, PROKOFEV1998253} as formulated for two-component hard-core bosons \cite{PhysRevA.81.053622}.  We adapted those techniques to incorporate the hard-core interspecies interactions from Eq.~(\ref{eq:ham_bh}). 
Related Monte-Carlo studies were performed by Kaurov, Kuklov, and Meyerovich~\cite{kaurov2005}, who used a discrete-time Monte Carlo approach with the worm algorithm to simulate a two-color J-current model  on a two-dimensional square lattice, effectively modeling hard-core two-component boson systems. However, the accuracy of their results were limited by systematic errors arising from time discretization and temperature dependence.

The superfluid stiffness for each component is determined using the winding number~\cite{Ceperley:1989hb}, expressed as $\rho_{ij} = \langle \mathbf{W_i}\cdot  \mathbf{W_j}\rangle / dL^{d-2}\beta$, where 
the components of the vector $\mathbf{W_i}$ count how many times paths of particles to type $i$ wind about each direction in the periodic unit cell.  Here $d=2$ is the dimensionality of the system, $L$ the linear system size, and $\beta$ is the inverse temperature.
The covariance matrix for the winding numbers is diagonalized by considering symmetric and antisymmetric combinations, and we extract $\rho_{\pm}=\langle \mathbf{W_{\pm}^2} \rangle / dL^{d-2}\beta$, where $\mathbf{W_{\pm}} = \mathbf{W_1} \pm \mathbf{W_2}$.  As already discussed, these relate to mass and spin transport.  The drag coefficient is $\kappa=(\rho_+-\rho_-)/(\rho_++\rho_-)$.  As with the calculations in Sec.~\ref{MPS}, here we work in the grand-canonical ensemble, controlling particle numbers by adjusting the chemical potentials.


\section{Results}

\begin{figure}
    \centering
    \includegraphics[width=\columnwidth]{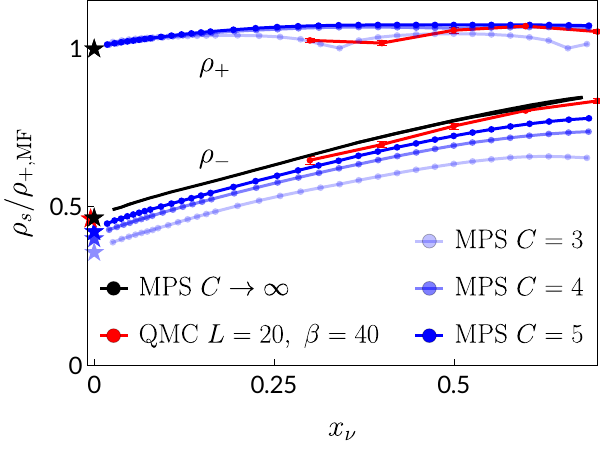}
    \caption{Superfluid fractions, $\rho_s$, in both the density ($\rho_+$, top) and spin ($\rho_-$, bottom) channels as a function of total hole density $x_\nu=1-n_\uparrow-n_\downarrow$. { Superfluid densities are normalized by the mean-field SF density $\rho_{+,{\rm MF}}=x_\nu(1-x_\nu)$.} Blue dots show results of the iMPS calculations on infinite cylinders of circumferences $C=3,4,5$, and the black line shows an extrapolation $C\to\infty$ (see main text). Stars denote the analytic results in the 
    {
    limit of vanishing hole density { for finite cylinders (blue), a $20\times20$ torus (red) and in the thermodynamic limit (black)}.}
    Note that the analytic calculation predicts no circumference dependence for $\rho_+$ as $x_\nu\to 0$, hence only one star is shown.
    {
    In all cases we consider a square lattice.  The dip in $\rho_+$ at $x_v=1/3$ for the $C=3$ cylinder is related to commensurability.} 
    Red dots show quantum Monte Carlo results on a $20\times20$ torus at inverse temperature $\beta=40/t$, which agrees quite well with the iMPS extrapolation for moderate dopings. 
    }
    \label{fig:sf_densities}
\end{figure}

Figure~\ref{fig:sf_densities}  shows the superfluid fractions $\rho_{\pm}$, { normalized by the mean-field superfluid density $\rho_{+,{\rm MF}}=x_\nu(1-x_\nu)$,} as a function of total hole density 
for {
iMPS calculations on} cylinders of circumference $C=3,4,5$ (blue).
Stars of the associated color show the perturbative results at infinitesimal doping, which are in close agreement with the iMPS calculations on every cylindrical geometry shown. Black stars show the perturbative result in the full 2D limit. The solid black curve shows the result of an extrapolation of {
the numerical} $\rho_-(C)$, where we assume the form $\rho_-(C)=\rho_-+\alpha/C^2$. We do not attempt a similar extrapolation of $\rho_+(C)$, as the $C=4$ and $C=5$ data are nearly indistinguishable.   { Red dots show the result of our quantum Monte Carlo calculations on a $20\times 20$ torus at inverse temperature $\beta=40/t$, over the parameter range where finite size/temperature effects are reasonably small (see Appendix~\ref{convergence}).}  

For all $x_v$, we find that $\rho_+$ remains 
close to the mean-field prediction 
$\rho_{+,{\rm MF}}=x_v(1-x_v)$. Indeed, $\rho_+$ in this model is simply the superfluid density of hard-core bosons, which is known to deviate only small amounts from $\rho_{+,{\rm MF}}$~\cite{Altman_2003}. The spin channel, by contrast, shows a significantly diminished stiffness relative to the mean-field prediction. This deviation increases in the dilute-hole limit, where the kinetics are more constrained.  Conversely, when the hole density is large, $x_\nu\to 1$, we expect $\rho_-\to\rho_{+,{\rm MF}}$ as there the interactions are negligible.  The kinetic effects of interactions are enhanced on smaller cylinders, and $\rho_-$ monotonically increases with $C$.

When $C=3$ we  see features in $\rho_+$  near $x_v\approx 1/3$ and $2/3$, which arise due to proximity to density-wave states.  In particular,  one can create a Mott insulator if  the hopping along the cylinder, $t_x$, is made much smaller than the hopping around the circumference, $t_y$.  Similar physics occurs on the 2-leg ladder at $x_v\approx 1/2$ \cite{crepin2011}.  We do not, however, see such features in the larger cylinders.  For $x_\nu<0.5$ we see a small deviation between the iMPS and Monte Carlo calculations of $\rho_+$, which we attribute to finite size/temperature.

Similarly, when extrapolated to $C=\infty$, the iMPS calculation of $\rho_-$ agrees well with the Monte Carlo calculation.  Moreover, as $x_\nu\to0$ we see excellent agreement with the analytic calculations.





\begin{figure}
    \centering
    \includegraphics[width=\columnwidth]{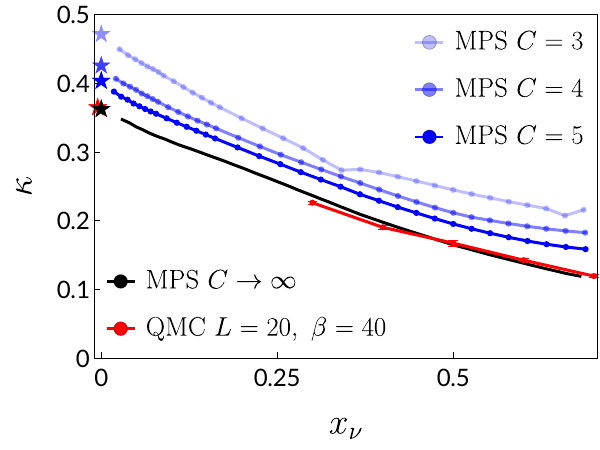}
    \caption{Superfluid drag coefficient, $\kappa=(\rho_+-\rho_-)/(\rho_++\rho_-)$, as a function of hole density $x_\nu$. Blue dots show results of the iMPS calculations on infinite cylinders of circumferences $C=3,4,5$, and the black line shows {
    the extrapolation of iMPS results as $C\to\infty$.} {
    Stars denote the analytic results on finite cylinders (blue), a $20\times20$ torus (red) and in the thermodynamic limit (black).}
    Red dots show quantum Monte Carlo results on a $20\times20$ torus at inverse temperature $\beta=40/t$, which agrees very well with the $C\to\infty$ limit of the iMPS results.
    }
    \label{fig:dragcoef}
\end{figure}

In Fig.~\ref{fig:dragcoef} we show the superfluid drag coefficient, $\kappa=(\rho_+-\rho_-)/(\rho_++\rho_-)$, as a function of $x_\nu$. Again, we find excellent agreement between the numerical calculations on finite cylinders (colored dots) and the analytic result (stars). The black curve denotes an extrapolation in which we compute $\kappa$ using the extrapolated $\rho_-$ (see above) and $\rho_+(C=5)$. This extrapolation agrees very well with our Monte-Carlo results.
We take the agreement between the three methods as a strong indication of the reliability of our results. The {
extrapolated} drag coefficient monotonically increases as $x_\nu\to 0$, approaching its asymptotic value $1-2/\pi$.

One way of interpreting these results is that, 
{
in the presence of spin currents,}
mobile holes are dressed by a ``polaron" of flipped spins. This {
polaronic physics is only apparent in the presence of counterflow supercurrents:}
As evidenced by Eq.~(\ref{eq:rotatedH}), the {
current-free} ground state in the rotated basis is simply the fully-polarized ground state of hard-core $b_{+}$ bosons at filling fraction $1-x_\nu$.

\begin{figure}
    \centering
    \includegraphics[width=\columnwidth]{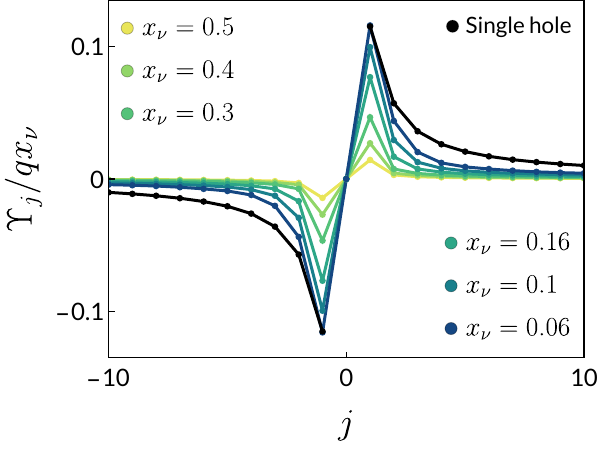}
    \caption{Spin-hole correlator $\Upsilon_j$ versus position $j$ at various hole densities $x_\nu$, determined from tensor network calculations on a $C=4$ circumference cylinder {
    (colored points).  The phase twist and the displacement $j$ are both taken in the long direction of the cylinder.} The correlation function is normalized by $q$ and $\langle P_0\rangle=x_\nu$. The analytic result for a single hole in two dimensions is shown in black. 
    }
    \label{fig:holecorr}
\end{figure}

{
To reveal the polaron, we introduce spin operators
$\tilde s_{xj}= (\tilde a_{j\uparrow}^\dagger \tilde a_{j_\downarrow} +\tilde a_{j\downarrow}^\dagger  \tilde a_{j_\uparrow} )/2$, $\tilde s_{yj}= i(\tilde a_{j\uparrow}^\dagger \tilde a_{j_\downarrow} -\tilde a_{j\downarrow}^\dagger \tilde a_{j_\uparrow} )/2$, and $\tilde s_{zj}=(\tilde a_{j\uparrow}^\dagger \tilde a_{j_\uparrow} -\tilde a_{j\downarrow}^\dagger \tilde a_{j_\downarrow} )/2$, 
which form the coefficients of the vector $\vec{\tilde s}_j$, defined in the rotated frame (see Eq.~(\ref{rot})).  We have a $xy$ ferromagnet, so the vector $\langle \vec{\tilde S}\rangle=\sum_j\langle \vec{\tilde s}_j\rangle/N_s$ has a non-zero expectation value in the $x-y$ plane.  Near a hole the spins will be twisted relative to this mean-field.  This twist can be quantified by $\Upsilon_{i-j}=\langle \vec{\tilde S}\rangle\times \langle P_i \vec{\tilde s}_j\rangle\cdot {\bf\hat{z}}$, where $P_i$ is the projector onto having a hole on site $i$. 
{ Equivalently, $\Upsilon_{j}\propto\langle \sin(\theta_j-\theta)\rangle$, where $\theta_j$ and $\theta$ are the in-plane angles of the spin a distance $j$ from the hole, and one very far from the hole.}
For the analytic wavefunction in Sec.~\ref{sec:pert} we have
\begin{equation}
\Upsilon_j=\frac{1}{2i} f_j^* f_0,
\end{equation}
where we have used that $f_j$ is purely imaginary.  Thus in the single hole limit, $|\Upsilon_j|^2$ is proportional to the probability that a flipped spin is separated from the hole by $j$ sites.

We can produce a more physical interpretation of $\Upsilon_j$ by returning to the lab frame (undoing the transformation in Eq.~(\ref{rot})).  There, on average, the in-plane spins rotate with wave-vector $q$.  This twist costs energy due to the misalignment of neighboring spins.  In our variational ansatz the energy is reduced by twisting the spins faster when they are near a hole and slower when they are farther away.  This spatial dependence of the pitch is encoded in $\Upsilon$.  One could directly measure this quantity by using the techniques from \cite{Parsons2016,Cheuk2016,Boll2016,Koepsell2019,Koepsell2021,bohrdt,qiao2025realizationdopedquantumantiferromagnet}. As a consequence of the arguments in Appendix~\ref{sec:pert_app}, $\Upsilon_{j}\sim \left({j_x q_x+j_y q_y}\right)/|j|^2$ for large $|j|$.

Figure~\ref{fig:holecorr} shows $\Upsilon_j$ calculated from the MPS calculations {
on a $C=4$ circumference cylinder.}
Here we take the twist and the displacement $j$ to both point along the $x$ axis. As can be seen, the spin disturbance is localized near the hole.  It becomes larger with smaller hole density, with the expected power law tails appearing as $x_v\to 0$.}


\section{Experimental Considerations}
As already explained, the most obvious platform for exploring this physics is atoms in optical lattices, and a detailed discussion of approaches to measuring superfluid density in single component systems can be found in \cite{kiely2022}.  There are additional approaches to detecting superfluid drag.  For example, as argued in \cite{kaurov2005,sellin2018}, superfluid drag has dramatic impact on the vortex structures.     

Previous experiments have explored aspects of superfluidity in two-component lattice bosons \cite{gadway2010}.  Those experiments were focused on the case where the two components have very different hopping matrix elements.  Variants of those techniques can measure drag effects in the SU(2) symmetric case, though there are subtleties about timescales \cite{carlini2021}.  
More recent explorations of counterflow superfluidity introduced techniques which can also be used to diagnose drag effects \cite{Zheng2025}.
Furthermore, drag can be extracted from the AC spin conductivity \cite{sekino2023}, or polaronic couplings to cavity modes \cite{pradhan2024}.

Using the equivalence between hard-core bosons and spin degrees of freedom, this model may also be realized via a mapping onto internal states of atoms and molecules~\cite{homeier2024,qiao2025realizationdopedquantumantiferromagnet} or the excited states of a transmon~\cite{Yanay2020}. In the former case, the states of $\uparrow$, $\downarrow$ and empty sites are encoded in three internal states that exhibit substantial dipole-dipole coupling (in neutral atoms, these could be Rydberg states). Using a stroboscopic driving scheme, one can engineer an effective antiferromagnetic bosonic $t-J$ model in which the couplings $t_{ij}\sim r_{ij}^{-3}$ and $J_{ij}\sim r_{ij}^{-6}$ are long-ranged. The model studied here is recovered in the limit $t\gg J$~\cite{harris2024kineticmagnetismstripeorder,qiao2025realizationdopedquantumantiferromagnet}, albeit with corrections due to long-range tunneling. In the case of transmons, one can map the hole state onto the computational $|1\rangle$ state of the transmon, and the spin states onto the $|0\rangle$ and $|2\rangle$ states. Introducing capacitive couplings, the $|0\rangle$ and $|2\rangle$ excitations can tunnel between transmons with a matrix element $g$. The hard-core boson model studied here is realized in the limit $g\ll\eta$ where $\eta$ is the transmon nonlinearity.

\section{Summary}
The most iconic problems in many-body physics involve the competition between strong local interactions, which tend to localize particles, and kinetic effects, which delocalize them.  Here we argue that superfluid drag in two-component lattice bosons provides an ideal setting for studying this physics.  Near the jamming limit of one particle per site, we find that the drag coefficient is of order unity: $\kappa\approx0.36$ on a 2D square lattice. By comparison, $\kappa$ vanishes in mean field theory. Such a large drag is a remarkable signature of strong correlations.  

We use three approaches to extract the drag coefficient.  The first is an analytic study of a single hole.  We calculate the exact quantum wavefunction in the presence of a small phase gradient.  We find a polaronic structure where the pseudospin twists faster near the hole.  From the energy of this state we extract the drag coefficient.  Our analytic calculation does not readily generalize to finite hole density, motivating a variational approach based upon matrix product states.  We use the VUMPS algorithm, and finite entanglement scaling, to extract the ground state energy in an infinite cylinder geometry, with circumference $C$.  Extrapolating to $C=\infty$ gives excellent agreement with our analytic calculation at vanishing hole density.  To develop further confidence in our extrapolation, we compared these results to quantum Monte-Carlo on large lattices with hundreds of sites.

We argue that this physics is experimentally accessible, using cold atoms in optical lattices, Rydberg atoms in microtraps, or transmon based quantum computers. 
{ Our model applies even if the  interactions are spin dependent, as long as  $t/U\ll1$.  We largely quoted results for two dimensional square lattices, but the} 
 calculations can readily be generalized to other geometries.

\section*{Acknowledgements}
We thank Anatoly Kuklov, Nikolay Prokof'ev, and Darren Pereira for discussions.
TGK acknowledges support from the National Science Foundation under grant PHY-2309135 to the Kavli Institute for Theoretical Physics (KITP), and
the
Gordon and Betty Moore Foundation through Grant GBMF8690 to the University of California, Santa Barbara.
CZ acknowledges support from the National Natural Science Foundation of China (NSFC) under Grant Nos. 12204173, 12275002, and the University Annual Scientific Research Plan of Anhui Province under Grant No 2022AH010013.
EJM acknowledges support from the National Science Foundation under grant  PHY-2409403.
\appendix

\section{Details of analytic calculation\label{sec:pert_app}}
Here we fill in the remaining steps of the calculation outlined in Sec.~\ref{sec:pert}. 
{ We first note that we only need to calculate $E(k,q)$ to quadratic order in $k$ and $q$.  Consequently we only need $f_s$ to linear order, and  on the right hand side of Eq.~(\ref{fs}) we  
take
\begin{equation}
\Gamma_s \approx i q_x \Pi_s^x +i q_y  \Pi_s^y 
\end{equation}
where
\begin{align}
\Pi_s^x&= t (\Lambda_{s_x+1,s_y}-\Lambda_{s_x-1,s_y})\\
\Pi_s^y&= t (\Lambda_{s_x,s_y+1}-\Lambda_{s_x,s_y+1}).
\end{align}
We can also set $k=q=0$ in our evaluation of $\Lambda_s$,
\begin{equation}
\Lambda_s \approx\frac{1}{2tN_s} \sum_{\vec p} \frac{e^{i\vec{p}\cdot \vec{s}}}{\cos p_x+\cos p_y-2}.
\end{equation}
In these linearized expressions, $\Lambda_{-s}=\Lambda_{s}$ and $\Gamma_{-s}=-\Gamma_s$.  Consequently $f_{-s}=-f_s$.  This allows us to write Eq.~(\ref{fs}) as
\begin{equation}\label{fs2}
f_s =
 (i q_x f_0-f_x)\Pi_s^x
+ (i q_y f_0-f_y)\Pi_s^y.
\end{equation}
Specializing to the cases $s=(1,0)$ and $s=(0,1)$ we find
\begin{align}
f_x &= i q_x f_0 \frac{\Pi^x_x}{1+\Pi^x_x}&
f_y &= i q_y f_0 \frac{\Pi^y_y}{1+\Pi^y_y}\label{fxy}
\end{align}
where $f_x=f_{(1,0)}$, $f_y=f_{(0,1)}$, $\Pi_x^x=\Pi^x_{(1,0)}$ and 
 $\Pi_y^x=\Pi^y_{(0,1)}$.  
 Inserting  Eq.~(\ref{fxy}) into Eq.~(\ref{fs2}) gives us the closed form expression
\begin{equation}
f_s= i q_x f_0 \frac{\Pi_s^x}{1+\Pi^x_x} + i q_y f_0 \frac{\Pi_s^y}{1+\Pi^y_y} 
\end{equation}
 which can be slightly simplified by noting that 
 rotational symmetry implies $\Pi_x^x=\Pi_y^y$.  We then use Eq.~(\ref{f0eqn}) to arrive at
$
E=\epsilon_0-2q^2 t\Pi_x^x/(1-\Pi_x^x).
$

To quadratic order, $\epsilon_0= t(-4+k^2+q^2)$, and hence
\begin{align}
E/t&=-4+k^2+\frac{1-\Pi^x_x}{1+\Pi^x_x} q^2.
\end{align}
Comparing to Eq.~(\ref{eq:pertenergy}),
we identify $\kappa=\Pi_x^x$ as the drag coefficient.  

To calculate $\kappa$ and the coefficients of the wavefunction, we take the thermodynamic limit and write
\begin{align}
\Pi^x_s &=\int \frac{dp_x\,dp_y}{(2\pi)^2}\frac{e^{i s_x p_x +is_y p_y} (e^{i p_x}-e^{-i p_x})}{2\cos p_x+2\cos p_y -4}.
\end{align}
We evaluate this integral by converting the $p_x$ integral into a contour integral with the substitution $z=e^{i p_x}$, yielding
\begin{align}
\Pi^x_s &=\int\frac{d p_y}{2\pi} 
e^{i s_y p_y}
\oint \frac{dz}{2\pi i} 
\frac{z^{s_x+1}-z^{s_x-1}}{z^2-2 w z+1}
\end{align}
where we have defined $w=2-\cos(p_y)$.  
By using the symmetries of the square lattice we can always take $s_x>0$, in which case 
the integrand's only  poles are at $z_{\pm}=w\pm\sqrt{w^2-1}=e^{\pm\chi}$, where $\cosh\chi=w$.  Only $z_-$ is inside the contour, and using the residue theorem
\begin{equation}\label{piy}
\Pi_s^x = \int\frac{d p_y}{2\pi} 
e^{i s_y p_y- s_x \chi}.
\end{equation}
This one dimensional integral can be efficiently calculated numerically.  For specific values of $s$ we can analytically calculate it.  For example
\begin{align}
\kappa&=\Pi_x^x=\int \frac{dp_y}{2\pi} \left(w-\sqrt{w^2-1}\right)\\
&=2-\int \frac{dp_y}{2\pi} \sqrt{w^2-1}\\
&=1-\frac{2}{\pi},
\end{align}
where the integral is performed by making the substitution $w=1+2\cos(2y)$.

A similar approach can be used to calculate the 
drag coefficient on a cylinder, where the integral over $p_y$ is replaced by a discrete sum.  In particular, on a cylinder of circumference $C$
\begin{equation}
\kappa =\frac{1}{C}\sum_{n=0}^{C-1} z_-(p_y=2\pi n/C).
\end{equation}
This expression was used to determine the stars in Fig.~(\ref{fig:dragcoef}).

For large $s_x$, the integrand in Eq.~(\ref{piy}) is dominated by $p_y$ near zero, where $\chi\approx |p_y|$, and hence
\begin{align}
\Pi^x_{s}&\approx \frac{1}{\pi} \frac{s_x}{s_x^2+s_y^2}.&(|s_x|\gg1)
\end{align}
}

\section{Convergence of Monte Carlo Calculation}\label{convergence}

\begin{figure}
    \centering
    \includegraphics[width=\columnwidth]{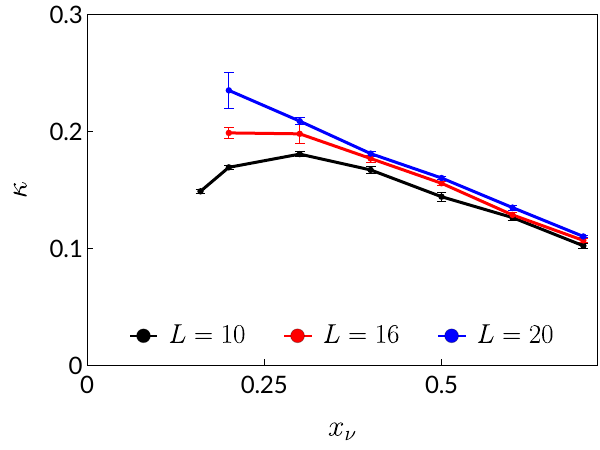}
    \caption{Superfluid drag coefficient $\kappa$ as a function of hole density $x_\nu$
    obtained from winding number statistics in quantum Monte Carlo (QMC) simulations on a square lattice for system sizes $L=10$ (black), $L=16$ (red), and $L=20$ (blue) at an inverse temperature $\beta=L$. 
    Error bars, if not shown, are within the symbol size.
    }
    \label{fig1}
\end{figure}

Here, we analyze both the finite-size and temperature dependence of $\kappa$, calculated with our Monte Carlo simulations.
Figure~\ref{fig1} shows $\kappa$ for square $L\times L$ lattices with 
%
 $L=10$, $16$, and $20$ at a inverse temperature $\beta=L$. As described in Sec.~\ref{mc}, the drag coefficient is calculated from winding number statistics.
In these calculations we find that $\kappa$ grows with $L$.  This is opposite to the $C$ dependence of the zero temperature iMPS calculations shown in Fig.~\ref{fig:dragcoef}. { The likely source of this behavior is the fact that we are simultaneously changing $L$ and $\beta$.  As we will demonstrate below, increasing temperature at fixed $L$ suppresses $\kappa$.} 
 Regardless, the $L$ and $\beta$ dependencies are largest at small hole densities.  For $L=\beta=20$ we see that the systematic errors from finite system size are of order the stochastic errors when  $x_\nu>0.3$.
 
\begin{figure}[tbph]
    \centering
    \includegraphics[width=\columnwidth]{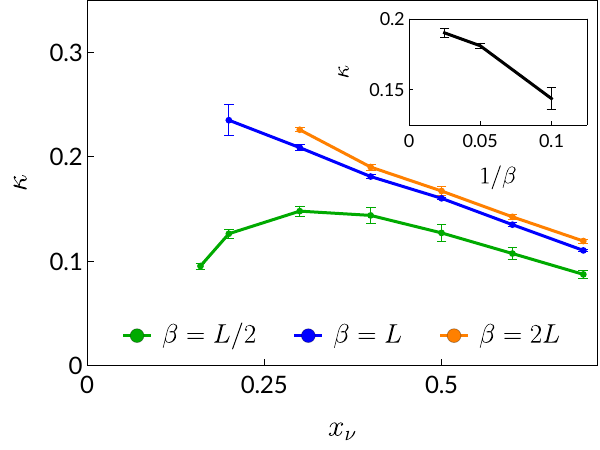}
    \caption{Superfluid drag coefficient $\kappa$ as a function of hole density $x_\nu$
    from quantum Monte Carlo (QMC) simulations on a square lattice with linear system size $L=20$. Data are shown for inverse temperatures $\beta=L/2$ (green), $\beta=L$ (blue), and $\beta=2L$ (orange). 
    Inset shows the superfluid drag coefficient $\kappa$ as a function of temperature $1/\beta$ for system size $L=20$ and hole density $x_\nu=0.4$. Error bars, if not visible, are within the symbol size.
    }
    \label{fig2}
\end{figure}

To 
help disentangle finite size and finite temperature effects, in Fig~\ref{fig2} we fix $L=20$ and calculate $\kappa$ for three different inverse temperatures: 
$\beta=L/2$, $\beta=L$, and $\beta=2L$.  Again, the temperature dependence is strongest at small $x_\nu$.  The drag coefficient grows with decreasing temperature, but appears to be a significant fraction of its zero-temperature value, even for $\beta=L/2$.  For $x_\nu>0.3$ the drag coefficient has largely saturated to its zero temperature value as long as $\beta>L$.  This behavior is better seen in the inset, which shows $\kappa$ as a function of temperature $1/\beta$ for system size $L=20$ and hole density $x_\nu=0.4$.

Due to computational constraints, we were unable to simulate system sizes larger than $L=20$ or inverse temperatures greater than $\beta=2L$.  The autocorrelation time in our simulations becomes larger at small $x_\nu$ due to kinetic constraints.  These correlations lead to larger statistical errors.  Similarly, increasing the system size leads to larger errors unless we greatly increase the number of Monte-Carlo samples.  


\bibliographystyle{apsrev4-2}

\begin{thebibliography}{51}%
\makeatletter
\providecommand \@ifxundefined [1]{%
 \@ifx{#1\undefined}
}%
\providecommand \@ifnum [1]{%
 \ifnum #1\expandafter \@firstoftwo
 \else \expandafter \@secondoftwo
 \fi
}%
\providecommand \@ifx [1]{%
 \ifx #1\expandafter \@firstoftwo
 \else \expandafter \@secondoftwo
 \fi
}%
\providecommand \natexlab [1]{#1}%
\providecommand \enquote  [1]{``#1''}%
\providecommand \bibnamefont  [1]{#1}%
\providecommand \bibfnamefont [1]{#1}%
\providecommand \citenamefont [1]{#1}%
\providecommand \href@noop [0]{\@secondoftwo}%
\providecommand \href [0]{\begingroup \@sanitize@url \@href}%
\providecommand \@href[1]{\@@startlink{#1}\@@href}%
\providecommand \@@href[1]{\endgroup#1\@@endlink}%
\providecommand \@sanitize@url [0]{\catcode `\\12\catcode `\$12\catcode
  `\&12\catcode `\#12\catcode `\^12\catcode `\_12\catcode `\%12\relax}%
\providecommand \@@startlink[1]{}%
\providecommand \@@endlink[0]{}%
\providecommand \url  [0]{\begingroup\@sanitize@url \@url }%
\providecommand \@url [1]{\endgroup\@href {#1}{\urlprefix }}%
\providecommand \urlprefix  [0]{URL }%
\providecommand \Eprint [0]{\href }%
\providecommand \doibase [0]{https://doi.org/}%
\providecommand \selectlanguage [0]{\@gobble}%
\providecommand \bibinfo  [0]{\@secondoftwo}%
\providecommand \bibfield  [0]{\@secondoftwo}%
\providecommand \translation [1]{[#1]}%
\providecommand \BibitemOpen [0]{}%
\providecommand \bibitemStop [0]{}%
\providecommand \bibitemNoStop [0]{.\EOS\space}%
\providecommand \EOS [0]{\spacefactor3000\relax}%
\providecommand \BibitemShut  [1]{\csname bibitem#1\endcsname}%
\let\auto@bib@innerbib\@empty
\bibitem [{\citenamefont {{Andreev}}\ and\ \citenamefont
  {{Bashkin}}(1975)}]{andreevbashkin}%
  \BibitemOpen
  \bibfield  {author} {\bibinfo {author} {\bibfnamefont {A.~F.}\ \bibnamefont
  {{Andreev}}}\ and\ \bibinfo {author} {\bibfnamefont {E.~P.}\ \bibnamefont
  {{Bashkin}}},\ }\href@noop {} {\bibfield  {journal} {\bibinfo  {journal}
  {Soviet Journal of Experimental and Theoretical Physics}\ }\textbf {\bibinfo
  {volume} {42}},\ \bibinfo {pages} {164} (\bibinfo {year} {1975})}\BibitemShut
  {NoStop}%
\bibitem [{\citenamefont {Fil}\ and\ \citenamefont
  {Shevchenko}(2005)}]{PhysRevA.72.013616}%
  \BibitemOpen
  \bibfield  {author} {\bibinfo {author} {\bibfnamefont {D.~V.}\ \bibnamefont
  {Fil}}\ and\ \bibinfo {author} {\bibfnamefont {S.~I.}\ \bibnamefont
  {Shevchenko}},\ }\href {https://doi.org/10.1103/PhysRevA.72.013616}
  {\bibfield  {journal} {\bibinfo  {journal} {Phys. Rev. A}\ }\textbf {\bibinfo
  {volume} {72}},\ \bibinfo {pages} {013616} (\bibinfo {year}
  {2005})}\BibitemShut {NoStop}%
\bibitem [{\citenamefont {Fil}\ and\ \citenamefont
  {Shevchenko}(2004)}]{Fil2004}%
  \BibitemOpen
  \bibfield  {author} {\bibinfo {author} {\bibfnamefont {D.~V.}\ \bibnamefont
  {Fil}}\ and\ \bibinfo {author} {\bibfnamefont {S.~I.}\ \bibnamefont
  {Shevchenko}},\ }\href {https://doi.org/10.1063/1.1808194} {\bibfield
  {journal} {\bibinfo  {journal} {Low Temperature Physics}\ }\textbf {\bibinfo
  {volume} {30}},\ \bibinfo {pages} {770–777} (\bibinfo {year}
  {2004})}\BibitemShut {NoStop}%
\bibitem [{\citenamefont {Nespolo}\ \emph {et~al.}(2017)\citenamefont
  {Nespolo}, \citenamefont {Astrakharchik},\ and\ \citenamefont
  {Recati}}]{Nespolo_2017}%
  \BibitemOpen
  \bibfield  {author} {\bibinfo {author} {\bibfnamefont {J.}~\bibnamefont
  {Nespolo}}, \bibinfo {author} {\bibfnamefont {G.~E.}\ \bibnamefont
  {Astrakharchik}},\ and\ \bibinfo {author} {\bibfnamefont {A.}~\bibnamefont
  {Recati}},\ }\href {https://doi.org/10.1088/1367-2630/aa93a0} {\bibfield
  {journal} {\bibinfo  {journal} {New Journal of Physics}\ }\textbf {\bibinfo
  {volume} {19}},\ \bibinfo {pages} {125005} (\bibinfo {year}
  {2017})}\BibitemShut {NoStop}%
\bibitem [{\citenamefont {Karle}\ \emph {et~al.}(2019)\citenamefont {Karle},
  \citenamefont {Defenu},\ and\ \citenamefont {Enss}}]{karle2019}%
  \BibitemOpen
  \bibfield  {author} {\bibinfo {author} {\bibfnamefont {V.}~\bibnamefont
  {Karle}}, \bibinfo {author} {\bibfnamefont {N.}~\bibnamefont {Defenu}},\ and\
  \bibinfo {author} {\bibfnamefont {T.}~\bibnamefont {Enss}},\ }\href
  {https://doi.org/10.1103/PhysRevA.99.063627} {\bibfield  {journal} {\bibinfo
  {journal} {Phys. Rev. A}\ }\textbf {\bibinfo {volume} {99}},\ \bibinfo
  {pages} {063627} (\bibinfo {year} {2019})}\BibitemShut {NoStop}%
\bibitem [{\citenamefont {Romito}\ \emph {et~al.}(2021)\citenamefont {Romito},
  \citenamefont {Lobo},\ and\ \citenamefont {Recati}}]{romito2021}%
  \BibitemOpen
  \bibfield  {author} {\bibinfo {author} {\bibfnamefont {D.}~\bibnamefont
  {Romito}}, \bibinfo {author} {\bibfnamefont {C.}~\bibnamefont {Lobo}},\ and\
  \bibinfo {author} {\bibfnamefont {A.}~\bibnamefont {Recati}},\ }\href
  {https://doi.org/10.1103/PhysRevResearch.3.023196} {\bibfield  {journal}
  {\bibinfo  {journal} {Phys. Rev. Res.}\ }\textbf {\bibinfo {volume} {3}},\
  \bibinfo {pages} {023196} (\bibinfo {year} {2021})}\BibitemShut {NoStop}%
\bibitem [{\citenamefont {Kaurov}\ \emph {et~al.}(2005)\citenamefont {Kaurov},
  \citenamefont {Kuklov},\ and\ \citenamefont {Meyerovich}}]{kaurov2005}%
  \BibitemOpen
  \bibfield  {author} {\bibinfo {author} {\bibfnamefont {V.~M.}\ \bibnamefont
  {Kaurov}}, \bibinfo {author} {\bibfnamefont {A.~B.}\ \bibnamefont {Kuklov}},\
  and\ \bibinfo {author} {\bibfnamefont {A.~E.}\ \bibnamefont {Meyerovich}},\
  }\href {https://doi.org/10.1103/PhysRevLett.95.090403} {\bibfield  {journal}
  {\bibinfo  {journal} {Phys. Rev. Lett.}\ }\textbf {\bibinfo {volume} {95}},\
  \bibinfo {pages} {090403} (\bibinfo {year} {2005})}\BibitemShut {NoStop}%
\bibitem [{\citenamefont {Sellin}\ and\ \citenamefont
  {Babaev}(2018)}]{sellin2018}%
  \BibitemOpen
  \bibfield  {author} {\bibinfo {author} {\bibfnamefont {K.}~\bibnamefont
  {Sellin}}\ and\ \bibinfo {author} {\bibfnamefont {E.}~\bibnamefont
  {Babaev}},\ }\href {https://doi.org/10.1103/PhysRevB.97.094517} {\bibfield
  {journal} {\bibinfo  {journal} {Phys. Rev. B}\ }\textbf {\bibinfo {volume}
  {97}},\ \bibinfo {pages} {094517} (\bibinfo {year} {2018})}\BibitemShut
  {NoStop}%
\bibitem [{\citenamefont {Kuklov}\ \emph
  {et~al.}(2004{\natexlab{a}})\citenamefont {Kuklov}, \citenamefont
  {Prokof'ev},\ and\ \citenamefont {Svistunov}}]{kuklov2004}%
  \BibitemOpen
  \bibfield  {author} {\bibinfo {author} {\bibfnamefont {A.}~\bibnamefont
  {Kuklov}}, \bibinfo {author} {\bibfnamefont {N.}~\bibnamefont {Prokof'ev}},\
  and\ \bibinfo {author} {\bibfnamefont {B.}~\bibnamefont {Svistunov}},\ }\href
  {https://doi.org/10.1103/PhysRevLett.92.030403} {\bibfield  {journal}
  {\bibinfo  {journal} {Phys. Rev. Lett.}\ }\textbf {\bibinfo {volume} {92}},\
  \bibinfo {pages} {030403} (\bibinfo {year} {2004}{\natexlab{a}})}\BibitemShut
  {NoStop}%
\bibitem [{\citenamefont {Kuklov}\ \emph
  {et~al.}(2004{\natexlab{b}})\citenamefont {Kuklov}, \citenamefont
  {Prokof'ev},\ and\ \citenamefont {Svistunov}}]{kuklov2004b}%
  \BibitemOpen
  \bibfield  {author} {\bibinfo {author} {\bibfnamefont {A.}~\bibnamefont
  {Kuklov}}, \bibinfo {author} {\bibfnamefont {N.}~\bibnamefont {Prokof'ev}},\
  and\ \bibinfo {author} {\bibfnamefont {B.}~\bibnamefont {Svistunov}},\ }\href
  {https://doi.org/10.1103/PhysRevLett.92.050402} {\bibfield  {journal}
  {\bibinfo  {journal} {Phys. Rev. Lett.}\ }\textbf {\bibinfo {volume} {92}},\
  \bibinfo {pages} {050402} (\bibinfo {year} {2004}{\natexlab{b}})}\BibitemShut
  {NoStop}%
\bibitem [{\citenamefont {Isacsson}\ \emph {et~al.}(2005)\citenamefont
  {Isacsson}, \citenamefont {Cha}, \citenamefont {Sengupta},\ and\
  \citenamefont {Girvin}}]{isacsson2005}%
  \BibitemOpen
  \bibfield  {author} {\bibinfo {author} {\bibfnamefont {A.}~\bibnamefont
  {Isacsson}}, \bibinfo {author} {\bibfnamefont {M.-C.}\ \bibnamefont {Cha}},
  \bibinfo {author} {\bibfnamefont {K.}~\bibnamefont {Sengupta}},\ and\
  \bibinfo {author} {\bibfnamefont {S.~M.}\ \bibnamefont {Girvin}},\ }\href
  {https://doi.org/10.1103/PhysRevB.72.184507} {\bibfield  {journal} {\bibinfo
  {journal} {Phys. Rev. B}\ }\textbf {\bibinfo {volume} {72}},\ \bibinfo
  {pages} {184507} (\bibinfo {year} {2005})}\BibitemShut {NoStop}%
\bibitem [{\citenamefont {Hu}\ \emph {et~al.}(2009)\citenamefont {Hu},
  \citenamefont {Mathey}, \citenamefont {Danshita}, \citenamefont {Tiesinga},
  \citenamefont {Williams},\ and\ \citenamefont {Clark}}]{hu2009}%
  \BibitemOpen
  \bibfield  {author} {\bibinfo {author} {\bibfnamefont {A.}~\bibnamefont
  {Hu}}, \bibinfo {author} {\bibfnamefont {L.}~\bibnamefont {Mathey}}, \bibinfo
  {author} {\bibfnamefont {I.}~\bibnamefont {Danshita}}, \bibinfo {author}
  {\bibfnamefont {E.}~\bibnamefont {Tiesinga}}, \bibinfo {author}
  {\bibfnamefont {C.~J.}\ \bibnamefont {Williams}},\ and\ \bibinfo {author}
  {\bibfnamefont {C.~W.}\ \bibnamefont {Clark}},\ }\href
  {https://doi.org/10.1103/PhysRevA.80.023619} {\bibfield  {journal} {\bibinfo
  {journal} {Phys. Rev. A}\ }\textbf {\bibinfo {volume} {80}},\ \bibinfo
  {pages} {023619} (\bibinfo {year} {2009})}\BibitemShut {NoStop}%
\bibitem [{\citenamefont {Kuklov}\ and\ \citenamefont
  {Svistunov}(2003)}]{kuklov2003}%
  \BibitemOpen
  \bibfield  {author} {\bibinfo {author} {\bibfnamefont {A.~B.}\ \bibnamefont
  {Kuklov}}\ and\ \bibinfo {author} {\bibfnamefont {B.~V.}\ \bibnamefont
  {Svistunov}},\ }\href {https://doi.org/10.1103/PhysRevLett.90.100401}
  {\bibfield  {journal} {\bibinfo  {journal} {Phys. Rev. Lett.}\ }\textbf
  {\bibinfo {volume} {90}},\ \bibinfo {pages} {100401} (\bibinfo {year}
  {2003})}\BibitemShut {NoStop}%
\bibitem [{\citenamefont {Altman}\ \emph {et~al.}(2003)\citenamefont {Altman},
  \citenamefont {Hofstetter}, \citenamefont {Demler},\ and\ \citenamefont
  {Lukin}}]{Altman_2003}%
  \BibitemOpen
  \bibfield  {author} {\bibinfo {author} {\bibfnamefont {E.}~\bibnamefont
  {Altman}}, \bibinfo {author} {\bibfnamefont {W.}~\bibnamefont {Hofstetter}},
  \bibinfo {author} {\bibfnamefont {E.}~\bibnamefont {Demler}},\ and\ \bibinfo
  {author} {\bibfnamefont {M.~D.}\ \bibnamefont {Lukin}},\ }\href
  {https://doi.org/10.1088/1367-2630/5/1/113} {\bibfield  {journal} {\bibinfo
  {journal} {New Journal of Physics}\ }\textbf {\bibinfo {volume} {5}},\
  \bibinfo {pages} {113} (\bibinfo {year} {2003})}\BibitemShut {NoStop}%
\bibitem [{\citenamefont {Hubener}\ \emph {et~al.}(2009)\citenamefont
  {Hubener}, \citenamefont {Snoek},\ and\ \citenamefont
  {Hofstetter}}]{hubener2009}%
  \BibitemOpen
  \bibfield  {author} {\bibinfo {author} {\bibfnamefont {A.}~\bibnamefont
  {Hubener}}, \bibinfo {author} {\bibfnamefont {M.}~\bibnamefont {Snoek}},\
  and\ \bibinfo {author} {\bibfnamefont {W.}~\bibnamefont {Hofstetter}},\
  }\href {https://doi.org/10.1103/PhysRevB.80.245109} {\bibfield  {journal}
  {\bibinfo  {journal} {Phys. Rev. B}\ }\textbf {\bibinfo {volume} {80}},\
  \bibinfo {pages} {245109} (\bibinfo {year} {2009})}\BibitemShut {NoStop}%
\bibitem [{\citenamefont {Nagaoka}(1966)}]{nagaoka1966}%
  \BibitemOpen
  \bibfield  {author} {\bibinfo {author} {\bibfnamefont {Y.}~\bibnamefont
  {Nagaoka}},\ }\href {https://doi.org/10.1103/PhysRev.147.392} {\bibfield
  {journal} {\bibinfo  {journal} {Phys. Rev.}\ }\textbf {\bibinfo {volume}
  {147}},\ \bibinfo {pages} {392} (\bibinfo {year} {1966})}\BibitemShut
  {NoStop}%
\bibitem [{\citenamefont {Eisenberg}\ and\ \citenamefont
  {Lieb}(2002)}]{eisenberg2002}%
  \BibitemOpen
  \bibfield  {author} {\bibinfo {author} {\bibfnamefont {E.}~\bibnamefont
  {Eisenberg}}\ and\ \bibinfo {author} {\bibfnamefont {E.~H.}\ \bibnamefont
  {Lieb}},\ }\href {https://doi.org/10.1103/PhysRevLett.89.220403} {\bibfield
  {journal} {\bibinfo  {journal} {Phys. Rev. Lett.}\ }\textbf {\bibinfo
  {volume} {89}},\ \bibinfo {pages} {220403} (\bibinfo {year}
  {2002})}\BibitemShut {NoStop}%
\bibitem [{\citenamefont {Parisi}\ \emph {et~al.}(2018)\citenamefont {Parisi},
  \citenamefont {Astrakharchik},\ and\ \citenamefont {Giorgini}}]{parisi}%
  \BibitemOpen
  \bibfield  {author} {\bibinfo {author} {\bibfnamefont {L.}~\bibnamefont
  {Parisi}}, \bibinfo {author} {\bibfnamefont {G.~E.}\ \bibnamefont
  {Astrakharchik}},\ and\ \bibinfo {author} {\bibfnamefont {S.}~\bibnamefont
  {Giorgini}},\ }\href {https://doi.org/10.1103/PhysRevLett.121.025302}
  {\bibfield  {journal} {\bibinfo  {journal} {Phys. Rev. Lett.}\ }\textbf
  {\bibinfo {volume} {121}},\ \bibinfo {pages} {025302} (\bibinfo {year}
  {2018})}\BibitemShut {NoStop}%
\bibitem [{\citenamefont {Dahl}\ \emph {et~al.}(2008)\citenamefont {Dahl},
  \citenamefont {Babaev},\ and\ \citenamefont {Sudb\o{}}}]{dahl}%
  \BibitemOpen
  \bibfield  {author} {\bibinfo {author} {\bibfnamefont {E.~K.}\ \bibnamefont
  {Dahl}}, \bibinfo {author} {\bibfnamefont {E.}~\bibnamefont {Babaev}},\ and\
  \bibinfo {author} {\bibfnamefont {A.}~\bibnamefont {Sudb\o{}}},\ }\href
  {https://doi.org/10.1103/PhysRevLett.101.255301} {\bibfield  {journal}
  {\bibinfo  {journal} {Phys. Rev. Lett.}\ }\textbf {\bibinfo {volume} {101}},\
  \bibinfo {pages} {255301} (\bibinfo {year} {2008})}\BibitemShut {NoStop}%
\bibitem [{\citenamefont {Parsons}\ \emph {et~al.}(2016)\citenamefont
  {Parsons}, \citenamefont {Mazurenko}, \citenamefont {Chiu}, \citenamefont
  {Ji}, \citenamefont {Greif},\ and\ \citenamefont {Greiner}}]{Parsons2016}%
  \BibitemOpen
  \bibfield  {author} {\bibinfo {author} {\bibfnamefont {M.~F.}\ \bibnamefont
  {Parsons}}, \bibinfo {author} {\bibfnamefont {A.}~\bibnamefont {Mazurenko}},
  \bibinfo {author} {\bibfnamefont {C.~S.}\ \bibnamefont {Chiu}}, \bibinfo
  {author} {\bibfnamefont {G.}~\bibnamefont {Ji}}, \bibinfo {author}
  {\bibfnamefont {D.}~\bibnamefont {Greif}},\ and\ \bibinfo {author}
  {\bibfnamefont {M.}~\bibnamefont {Greiner}},\ }\href
  {https://doi.org/10.1126/science.aag1430} {\bibfield  {journal} {\bibinfo
  {journal} {Science}\ }\textbf {\bibinfo {volume} {353}},\ \bibinfo {pages}
  {1253–1256} (\bibinfo {year} {2016})}\BibitemShut {NoStop}%
\bibitem [{\citenamefont {Cheuk}\ \emph {et~al.}(2016)\citenamefont {Cheuk},
  \citenamefont {Nichols}, \citenamefont {Lawrence}, \citenamefont {Okan},
  \citenamefont {Zhang}, \citenamefont {Khatami}, \citenamefont {Trivedi},
  \citenamefont {Paiva}, \citenamefont {Rigol},\ and\ \citenamefont
  {Zwierlein}}]{Cheuk2016}%
  \BibitemOpen
  \bibfield  {author} {\bibinfo {author} {\bibfnamefont {L.~W.}\ \bibnamefont
  {Cheuk}}, \bibinfo {author} {\bibfnamefont {M.~A.}\ \bibnamefont {Nichols}},
  \bibinfo {author} {\bibfnamefont {K.~R.}\ \bibnamefont {Lawrence}}, \bibinfo
  {author} {\bibfnamefont {M.}~\bibnamefont {Okan}}, \bibinfo {author}
  {\bibfnamefont {H.}~\bibnamefont {Zhang}}, \bibinfo {author} {\bibfnamefont
  {E.}~\bibnamefont {Khatami}}, \bibinfo {author} {\bibfnamefont
  {N.}~\bibnamefont {Trivedi}}, \bibinfo {author} {\bibfnamefont
  {T.}~\bibnamefont {Paiva}}, \bibinfo {author} {\bibfnamefont
  {M.}~\bibnamefont {Rigol}},\ and\ \bibinfo {author} {\bibfnamefont {M.~W.}\
  \bibnamefont {Zwierlein}},\ }\href {https://doi.org/10.1126/science.aag3349}
  {\bibfield  {journal} {\bibinfo  {journal} {Science}\ }\textbf {\bibinfo
  {volume} {353}},\ \bibinfo {pages} {1260–1264} (\bibinfo {year}
  {2016})}\BibitemShut {NoStop}%
\bibitem [{\citenamefont {Boll}\ \emph {et~al.}(2016)\citenamefont {Boll},
  \citenamefont {Hilker}, \citenamefont {Salomon}, \citenamefont {Omran},
  \citenamefont {Nespolo}, \citenamefont {Pollet}, \citenamefont {Bloch},\ and\
  \citenamefont {Gross}}]{Boll2016}%
  \BibitemOpen
  \bibfield  {author} {\bibinfo {author} {\bibfnamefont {M.}~\bibnamefont
  {Boll}}, \bibinfo {author} {\bibfnamefont {T.~A.}\ \bibnamefont {Hilker}},
  \bibinfo {author} {\bibfnamefont {G.}~\bibnamefont {Salomon}}, \bibinfo
  {author} {\bibfnamefont {A.}~\bibnamefont {Omran}}, \bibinfo {author}
  {\bibfnamefont {J.}~\bibnamefont {Nespolo}}, \bibinfo {author} {\bibfnamefont
  {L.}~\bibnamefont {Pollet}}, \bibinfo {author} {\bibfnamefont
  {I.}~\bibnamefont {Bloch}},\ and\ \bibinfo {author} {\bibfnamefont
  {C.}~\bibnamefont {Gross}},\ }\href {https://doi.org/10.1126/science.aag1635}
  {\bibfield  {journal} {\bibinfo  {journal} {Science}\ }\textbf {\bibinfo
  {volume} {353}},\ \bibinfo {pages} {1257–1260} (\bibinfo {year}
  {2016})}\BibitemShut {NoStop}%
\bibitem [{\citenamefont {Koepsell}\ \emph {et~al.}(2019)\citenamefont
  {Koepsell}, \citenamefont {Vijayan}, \citenamefont {Sompet}, \citenamefont
  {Grusdt}, \citenamefont {Hilker}, \citenamefont {Demler}, \citenamefont
  {Salomon}, \citenamefont {Bloch},\ and\ \citenamefont
  {Gross}}]{Koepsell2019}%
  \BibitemOpen
  \bibfield  {author} {\bibinfo {author} {\bibfnamefont {J.}~\bibnamefont
  {Koepsell}}, \bibinfo {author} {\bibfnamefont {J.}~\bibnamefont {Vijayan}},
  \bibinfo {author} {\bibfnamefont {P.}~\bibnamefont {Sompet}}, \bibinfo
  {author} {\bibfnamefont {F.}~\bibnamefont {Grusdt}}, \bibinfo {author}
  {\bibfnamefont {T.~A.}\ \bibnamefont {Hilker}}, \bibinfo {author}
  {\bibfnamefont {E.}~\bibnamefont {Demler}}, \bibinfo {author} {\bibfnamefont
  {G.}~\bibnamefont {Salomon}}, \bibinfo {author} {\bibfnamefont
  {I.}~\bibnamefont {Bloch}},\ and\ \bibinfo {author} {\bibfnamefont
  {C.}~\bibnamefont {Gross}},\ }\href
  {https://doi.org/10.1038/s41586-019-1463-1} {\bibfield  {journal} {\bibinfo
  {journal} {Nature}\ }\textbf {\bibinfo {volume} {572}},\ \bibinfo {pages}
  {358–362} (\bibinfo {year} {2019})}\BibitemShut {NoStop}%
\bibitem [{\citenamefont {Koepsell}\ \emph {et~al.}(2021)\citenamefont
  {Koepsell}, \citenamefont {Bourgund}, \citenamefont {Sompet}, \citenamefont
  {Hirthe}, \citenamefont {Bohrdt}, \citenamefont {Wang}, \citenamefont
  {Grusdt}, \citenamefont {Demler}, \citenamefont {Salomon}, \citenamefont
  {Gross},\ and\ \citenamefont {Bloch}}]{Koepsell2021}%
  \BibitemOpen
  \bibfield  {author} {\bibinfo {author} {\bibfnamefont {J.}~\bibnamefont
  {Koepsell}}, \bibinfo {author} {\bibfnamefont {D.}~\bibnamefont {Bourgund}},
  \bibinfo {author} {\bibfnamefont {P.}~\bibnamefont {Sompet}}, \bibinfo
  {author} {\bibfnamefont {S.}~\bibnamefont {Hirthe}}, \bibinfo {author}
  {\bibfnamefont {A.}~\bibnamefont {Bohrdt}}, \bibinfo {author} {\bibfnamefont
  {Y.}~\bibnamefont {Wang}}, \bibinfo {author} {\bibfnamefont {F.}~\bibnamefont
  {Grusdt}}, \bibinfo {author} {\bibfnamefont {E.}~\bibnamefont {Demler}},
  \bibinfo {author} {\bibfnamefont {G.}~\bibnamefont {Salomon}}, \bibinfo
  {author} {\bibfnamefont {C.}~\bibnamefont {Gross}},\ and\ \bibinfo {author}
  {\bibfnamefont {I.}~\bibnamefont {Bloch}},\ }\href
  {https://doi.org/10.1126/science.abe7165} {\bibfield  {journal} {\bibinfo
  {journal} {Science}\ }\textbf {\bibinfo {volume} {374}},\ \bibinfo {pages}
  {82–86} (\bibinfo {year} {2021})}\BibitemShut {NoStop}%
\bibitem [{\citenamefont {Bohrdt}\ \emph {et~al.}(2021)\citenamefont {Bohrdt},
  \citenamefont {Wang}, \citenamefont {Koepsell}, \citenamefont
  {K\'anasz-Nagy}, \citenamefont {Demler},\ and\ \citenamefont
  {Grusdt}}]{bohrdt}%
  \BibitemOpen
  \bibfield  {author} {\bibinfo {author} {\bibfnamefont {A.}~\bibnamefont
  {Bohrdt}}, \bibinfo {author} {\bibfnamefont {Y.}~\bibnamefont {Wang}},
  \bibinfo {author} {\bibfnamefont {J.}~\bibnamefont {Koepsell}}, \bibinfo
  {author} {\bibfnamefont {M.}~\bibnamefont {K\'anasz-Nagy}}, \bibinfo {author}
  {\bibfnamefont {E.}~\bibnamefont {Demler}},\ and\ \bibinfo {author}
  {\bibfnamefont {F.}~\bibnamefont {Grusdt}},\ }\href
  {https://doi.org/10.1103/PhysRevLett.126.026401} {\bibfield  {journal}
  {\bibinfo  {journal} {Phys. Rev. Lett.}\ }\textbf {\bibinfo {volume} {126}},\
  \bibinfo {pages} {026401} (\bibinfo {year} {2021})}\BibitemShut {NoStop}%
\bibitem [{\citenamefont {Linder}\ and\ \citenamefont
  {Sudb\o{}}(2009)}]{linder2009}%
  \BibitemOpen
  \bibfield  {author} {\bibinfo {author} {\bibfnamefont {J.}~\bibnamefont
  {Linder}}\ and\ \bibinfo {author} {\bibfnamefont {A.}~\bibnamefont
  {Sudb\o{}}},\ }\href {https://doi.org/10.1103/PhysRevA.79.063610} {\bibfield
  {journal} {\bibinfo  {journal} {Phys. Rev. A}\ }\textbf {\bibinfo {volume}
  {79}},\ \bibinfo {pages} {063610} (\bibinfo {year} {2009})}\BibitemShut
  {NoStop}%
\bibitem [{\citenamefont {Hofer}\ \emph {et~al.}(2012)\citenamefont {Hofer},
  \citenamefont {Bruder},\ and\ \citenamefont {Stojanovi\ifmmode~\acute{c}\else
  \'{c}\fi{}}}]{hofer2012}%
  \BibitemOpen
  \bibfield  {author} {\bibinfo {author} {\bibfnamefont {P.~P.}\ \bibnamefont
  {Hofer}}, \bibinfo {author} {\bibfnamefont {C.}~\bibnamefont {Bruder}},\ and\
  \bibinfo {author} {\bibfnamefont {V.~M.}\ \bibnamefont
  {Stojanovi\ifmmode~\acute{c}\else \'{c}\fi{}}},\ }\href
  {https://doi.org/10.1103/PhysRevA.86.033627} {\bibfield  {journal} {\bibinfo
  {journal} {Phys. Rev. A}\ }\textbf {\bibinfo {volume} {86}},\ \bibinfo
  {pages} {033627} (\bibinfo {year} {2012})}\BibitemShut {NoStop}%
\bibitem [{\citenamefont {Hartman}\ \emph {et~al.}(2018)\citenamefont
  {Hartman}, \citenamefont {Erlandsen},\ and\ \citenamefont
  {Sudb\o{}}}]{hartman2018}%
  \BibitemOpen
  \bibfield  {author} {\bibinfo {author} {\bibfnamefont {S.}~\bibnamefont
  {Hartman}}, \bibinfo {author} {\bibfnamefont {E.}~\bibnamefont {Erlandsen}},\
  and\ \bibinfo {author} {\bibfnamefont {A.}~\bibnamefont {Sudb\o{}}},\ }\href
  {https://doi.org/10.1103/PhysRevB.98.024512} {\bibfield  {journal} {\bibinfo
  {journal} {Phys. Rev. B}\ }\textbf {\bibinfo {volume} {98}},\ \bibinfo
  {pages} {024512} (\bibinfo {year} {2018})}\BibitemShut {NoStop}%
\bibitem [{\citenamefont {Colussi}\ \emph {et~al.}(2022)\citenamefont
  {Colussi}, \citenamefont {Caleffi}, \citenamefont {Menotti},\ and\
  \citenamefont {Recati}}]{colussi2022}%
  \BibitemOpen
  \bibfield  {author} {\bibinfo {author} {\bibfnamefont {V.~E.}\ \bibnamefont
  {Colussi}}, \bibinfo {author} {\bibfnamefont {F.}~\bibnamefont {Caleffi}},
  \bibinfo {author} {\bibfnamefont {C.}~\bibnamefont {Menotti}},\ and\ \bibinfo
  {author} {\bibfnamefont {A.}~\bibnamefont {Recati}},\ }\href
  {https://doi.org/10.21468/SciPostPhys.12.3.111} {\bibfield  {journal}
  {\bibinfo  {journal} {SciPost Phys.}\ }\textbf {\bibinfo {volume} {12}},\
  \bibinfo {pages} {111} (\bibinfo {year} {2022})}\BibitemShut {NoStop}%
\bibitem [{\citenamefont {Contessi}\ \emph {et~al.}(2021)\citenamefont
  {Contessi}, \citenamefont {Romito}, \citenamefont {Rizzi},\ and\
  \citenamefont {Recati}}]{contessi2021}%
  \BibitemOpen
  \bibfield  {author} {\bibinfo {author} {\bibfnamefont {D.}~\bibnamefont
  {Contessi}}, \bibinfo {author} {\bibfnamefont {D.}~\bibnamefont {Romito}},
  \bibinfo {author} {\bibfnamefont {M.}~\bibnamefont {Rizzi}},\ and\ \bibinfo
  {author} {\bibfnamefont {A.}~\bibnamefont {Recati}},\ }\href
  {https://doi.org/10.1103/PhysRevResearch.3.L022017} {\bibfield  {journal}
  {\bibinfo  {journal} {Phys. Rev. Res.}\ }\textbf {\bibinfo {volume} {3}},\
  \bibinfo {pages} {L022017} (\bibinfo {year} {2021})}\BibitemShut {NoStop}%
\bibitem [{\citenamefont {Gr\'emaud}\ and\ \citenamefont
  {Batrouni}(2021)}]{gremaud2021}%
  \BibitemOpen
  \bibfield  {author} {\bibinfo {author} {\bibfnamefont {B.}~\bibnamefont
  {Gr\'emaud}}\ and\ \bibinfo {author} {\bibfnamefont {G.~G.}\ \bibnamefont
  {Batrouni}},\ }\href {https://doi.org/10.1103/PhysRevLett.127.025301}
  {\bibfield  {journal} {\bibinfo  {journal} {Phys. Rev. Lett.}\ }\textbf
  {\bibinfo {volume} {127}},\ \bibinfo {pages} {025301} (\bibinfo {year}
  {2021})}\BibitemShut {NoStop}%
\bibitem [{\citenamefont {Kiely}\ and\ \citenamefont
  {Mueller}(2022{\natexlab{a}})}]{kiely2022}%
  \BibitemOpen
  \bibfield  {author} {\bibinfo {author} {\bibfnamefont {T.~G.}\ \bibnamefont
  {Kiely}}\ and\ \bibinfo {author} {\bibfnamefont {E.~J.}\ \bibnamefont
  {Mueller}},\ }\href {https://doi.org/10.1103/PhysRevB.105.134502} {\bibfield
  {journal} {\bibinfo  {journal} {Phys. Rev. B}\ }\textbf {\bibinfo {volume}
  {105}},\ \bibinfo {pages} {134502} (\bibinfo {year}
  {2022}{\natexlab{a}})}\BibitemShut {NoStop}%
\bibitem [{\citenamefont {Schollw\"{o}ck}(2011)}]{Schollwck2011}%
  \BibitemOpen
  \bibfield  {author} {\bibinfo {author} {\bibfnamefont {U.}~\bibnamefont
  {Schollw\"{o}ck}},\ }\href {https://doi.org/10.1016/j.aop.2010.09.012}
  {\bibfield  {journal} {\bibinfo  {journal} {Annals of Physics}\ }\textbf
  {\bibinfo {volume} {326}},\ \bibinfo {pages} {96–192} (\bibinfo {year}
  {2011})}\BibitemShut {NoStop}%
\bibitem [{\citenamefont {Liang}\ and\ \citenamefont {Pang}(1994)}]{liang1994}%
  \BibitemOpen
  \bibfield  {author} {\bibinfo {author} {\bibfnamefont {S.}~\bibnamefont
  {Liang}}\ and\ \bibinfo {author} {\bibfnamefont {H.}~\bibnamefont {Pang}},\
  }\href {https://doi.org/10.1103/PhysRevB.49.9214} {\bibfield  {journal}
  {\bibinfo  {journal} {Phys. Rev. B}\ }\textbf {\bibinfo {volume} {49}},\
  \bibinfo {pages} {9214} (\bibinfo {year} {1994})}\BibitemShut {NoStop}%
\bibitem [{\citenamefont {Zauner-Stauber}\ \emph {et~al.}(2018)\citenamefont
  {Zauner-Stauber}, \citenamefont {Vanderstraeten}, \citenamefont {Fishman},
  \citenamefont {Verstraete},\ and\ \citenamefont {Haegeman}}]{VUMPS}%
  \BibitemOpen
  \bibfield  {author} {\bibinfo {author} {\bibfnamefont {V.}~\bibnamefont
  {Zauner-Stauber}}, \bibinfo {author} {\bibfnamefont {L.}~\bibnamefont
  {Vanderstraeten}}, \bibinfo {author} {\bibfnamefont {M.~T.}\ \bibnamefont
  {Fishman}}, \bibinfo {author} {\bibfnamefont {F.}~\bibnamefont
  {Verstraete}},\ and\ \bibinfo {author} {\bibfnamefont {J.}~\bibnamefont
  {Haegeman}},\ }\href {https://doi.org/10.1103/PhysRevB.97.045145} {\bibfield
  {journal} {\bibinfo  {journal} {Phys. Rev. B}\ }\textbf {\bibinfo {volume}
  {97}},\ \bibinfo {pages} {045145} (\bibinfo {year} {2018})}\BibitemShut
  {NoStop}%
\bibitem [{\citenamefont {Kiely}\ and\ \citenamefont
  {Mueller}(2022{\natexlab{b}})}]{kiely2022b}%
  \BibitemOpen
  \bibfield  {author} {\bibinfo {author} {\bibfnamefont {T.~G.}\ \bibnamefont
  {Kiely}}\ and\ \bibinfo {author} {\bibfnamefont {E.~J.}\ \bibnamefont
  {Mueller}},\ }\href {https://doi.org/10.1103/PhysRevB.106.235126} {\bibfield
  {journal} {\bibinfo  {journal} {Phys. Rev. B}\ }\textbf {\bibinfo {volume}
  {106}},\ \bibinfo {pages} {235126} (\bibinfo {year}
  {2022}{\natexlab{b}})}\BibitemShut {NoStop}%
\bibitem [{\citenamefont {Pollmann}\ \emph {et~al.}(2009)\citenamefont
  {Pollmann}, \citenamefont {Mukerjee}, \citenamefont {Turner},\ and\
  \citenamefont {Moore}}]{pollmann2009}%
  \BibitemOpen
  \bibfield  {author} {\bibinfo {author} {\bibfnamefont {F.}~\bibnamefont
  {Pollmann}}, \bibinfo {author} {\bibfnamefont {S.}~\bibnamefont {Mukerjee}},
  \bibinfo {author} {\bibfnamefont {A.~M.}\ \bibnamefont {Turner}},\ and\
  \bibinfo {author} {\bibfnamefont {J.~E.}\ \bibnamefont {Moore}},\ }\href
  {https://doi.org/10.1103/PhysRevLett.102.255701} {\bibfield  {journal}
  {\bibinfo  {journal} {Phys. Rev. Lett.}\ }\textbf {\bibinfo {volume} {102}},\
  \bibinfo {pages} {255701} (\bibinfo {year} {2009})}\BibitemShut {NoStop}%
\bibitem [{\citenamefont {Prokof'ev}\ \emph
  {et~al.}(1998{\natexlab{a}})\citenamefont {Prokof'ev}, \citenamefont
  {Svistunov},\ and\ \citenamefont {Tupitsyn}}]{PROKOFEV1998}%
  \BibitemOpen
  \bibfield  {author} {\bibinfo {author} {\bibfnamefont {N.}~\bibnamefont
  {Prokof'ev}}, \bibinfo {author} {\bibfnamefont {B.}~\bibnamefont
  {Svistunov}},\ and\ \bibinfo {author} {\bibfnamefont {I.}~\bibnamefont
  {Tupitsyn}},\ }\href {https://doi.org/10.1134/1.558661} {\bibfield  {journal}
  {\bibinfo  {journal} {Journal of Experimental and Theoretical Physics}\
  }\textbf {\bibinfo {volume} {87}},\ \bibinfo {pages} {310} (\bibinfo {year}
  {1998}{\natexlab{a}})}\BibitemShut {NoStop}%
\bibitem [{\citenamefont {Prokof'ev}\ \emph
  {et~al.}(1998{\natexlab{b}})\citenamefont {Prokof'ev}, \citenamefont
  {Svistunov},\ and\ \citenamefont {Tupitsyn}}]{PROKOFEV1998253}%
  \BibitemOpen
  \bibfield  {author} {\bibinfo {author} {\bibfnamefont {N.}~\bibnamefont
  {Prokof'ev}}, \bibinfo {author} {\bibfnamefont {B.}~\bibnamefont
  {Svistunov}},\ and\ \bibinfo {author} {\bibfnamefont {I.}~\bibnamefont
  {Tupitsyn}},\ }\href
  {https://doi.org/https://doi.org/10.1016/S0375-9601(97)00957-2} {\bibfield
  {journal} {\bibinfo  {journal} {Physics Letters A}\ }\textbf {\bibinfo
  {volume} {238}},\ \bibinfo {pages} {253} (\bibinfo {year}
  {1998}{\natexlab{b}})}\BibitemShut {NoStop}%
\bibitem [{\citenamefont {Capogrosso-Sansone}\ \emph
  {et~al.}(2010)\citenamefont {Capogrosso-Sansone}, \citenamefont {S\"oyler},
  \citenamefont {Prokof'ev},\ and\ \citenamefont
  {Svistunov}}]{PhysRevA.81.053622}%
  \BibitemOpen
  \bibfield  {author} {\bibinfo {author} {\bibfnamefont {B.}~\bibnamefont
  {Capogrosso-Sansone}}, \bibinfo {author} {\bibfnamefont {G.}~\bibnamefont
  {S\"oyler}}, \bibinfo {author} {\bibfnamefont {N.~V.}\ \bibnamefont
  {Prokof'ev}},\ and\ \bibinfo {author} {\bibfnamefont {B.~V.}\ \bibnamefont
  {Svistunov}},\ }\href {https://doi.org/10.1103/PhysRevA.81.053622} {\bibfield
   {journal} {\bibinfo  {journal} {Phys. Rev. A}\ }\textbf {\bibinfo {volume}
  {81}},\ \bibinfo {pages} {053622} (\bibinfo {year} {2010})}\BibitemShut
  {NoStop}%
\bibitem [{\citenamefont {Ceperley}\ and\ \citenamefont
  {Pollock}(1989)}]{Ceperley:1989hb}%
  \BibitemOpen
  \bibfield  {author} {\bibinfo {author} {\bibfnamefont {D.~M.}\ \bibnamefont
  {Ceperley}}\ and\ \bibinfo {author} {\bibfnamefont {E.~L.}\ \bibnamefont
  {Pollock}},\ }\href {https://doi.org/10.1103/PhysRevB.39.2084} {\bibfield
  {journal} {\bibinfo  {journal} {Phys. Rev. B}\ }\textbf {\bibinfo {volume}
  {39}},\ \bibinfo {pages} {2084} (\bibinfo {year} {1989})}\BibitemShut
  {NoStop}%
\bibitem [{\citenamefont {Cr\'epin}\ \emph {et~al.}(2011)\citenamefont
  {Cr\'epin}, \citenamefont {Laflorencie}, \citenamefont {Roux},\ and\
  \citenamefont {Simon}}]{crepin2011}%
  \BibitemOpen
  \bibfield  {author} {\bibinfo {author} {\bibfnamefont {F.~m.~c.}\
  \bibnamefont {Cr\'epin}}, \bibinfo {author} {\bibfnamefont {N.}~\bibnamefont
  {Laflorencie}}, \bibinfo {author} {\bibfnamefont {G.}~\bibnamefont {Roux}},\
  and\ \bibinfo {author} {\bibfnamefont {P.}~\bibnamefont {Simon}},\ }\href
  {https://doi.org/10.1103/PhysRevB.84.054517} {\bibfield  {journal} {\bibinfo
  {journal} {Phys. Rev. B}\ }\textbf {\bibinfo {volume} {84}},\ \bibinfo
  {pages} {054517} (\bibinfo {year} {2011})}\BibitemShut {NoStop}%
\bibitem [{\citenamefont {Qiao}\ \emph {et~al.}(2025)\citenamefont {Qiao},
  \citenamefont {Emperauger}, \citenamefont {Chen}, \citenamefont {Homeier},
  \citenamefont {Hollerith}, \citenamefont {Bornet}, \citenamefont {Martin},
  \citenamefont {Gély}, \citenamefont {Klein}, \citenamefont {Barredo},
  \citenamefont {Geier}, \citenamefont {Chiu}, \citenamefont {Grusdt},
  \citenamefont {Bohrdt}, \citenamefont {Lahaye},\ and\ \citenamefont
  {Browaeys}}]{qiao2025realizationdopedquantumantiferromagnet}%
  \BibitemOpen
  \bibfield  {author} {\bibinfo {author} {\bibfnamefont {M.}~\bibnamefont
  {Qiao}}, \bibinfo {author} {\bibfnamefont {G.}~\bibnamefont {Emperauger}},
  \bibinfo {author} {\bibfnamefont {C.}~\bibnamefont {Chen}}, \bibinfo {author}
  {\bibfnamefont {L.}~\bibnamefont {Homeier}}, \bibinfo {author} {\bibfnamefont
  {S.}~\bibnamefont {Hollerith}}, \bibinfo {author} {\bibfnamefont
  {G.}~\bibnamefont {Bornet}}, \bibinfo {author} {\bibfnamefont
  {R.}~\bibnamefont {Martin}}, \bibinfo {author} {\bibfnamefont
  {B.}~\bibnamefont {Gély}}, \bibinfo {author} {\bibfnamefont
  {L.}~\bibnamefont {Klein}}, \bibinfo {author} {\bibfnamefont
  {D.}~\bibnamefont {Barredo}}, \bibinfo {author} {\bibfnamefont
  {S.}~\bibnamefont {Geier}}, \bibinfo {author} {\bibfnamefont {N.-C.}\
  \bibnamefont {Chiu}}, \bibinfo {author} {\bibfnamefont {F.}~\bibnamefont
  {Grusdt}}, \bibinfo {author} {\bibfnamefont {A.}~\bibnamefont {Bohrdt}},
  \bibinfo {author} {\bibfnamefont {T.}~\bibnamefont {Lahaye}},\ and\ \bibinfo
  {author} {\bibfnamefont {A.}~\bibnamefont {Browaeys}},\ }\href
  {https://arxiv.org/abs/2501.08233} {\bibinfo {title} {Realization of a doped
  quantum antiferromagnet with dipolar tunnelings in a rydberg tweezer array}}
  (\bibinfo {year} {2025}),\ \Eprint {https://arxiv.org/abs/2501.08233}
  {arXiv:2501.08233 [quant-ph]} \BibitemShut {NoStop}%
\bibitem [{\citenamefont {Gadway}\ \emph {et~al.}(2010)\citenamefont {Gadway},
  \citenamefont {Pertot}, \citenamefont {Reimann},\ and\ \citenamefont
  {Schneble}}]{gadway2010}%
  \BibitemOpen
  \bibfield  {author} {\bibinfo {author} {\bibfnamefont {B.}~\bibnamefont
  {Gadway}}, \bibinfo {author} {\bibfnamefont {D.}~\bibnamefont {Pertot}},
  \bibinfo {author} {\bibfnamefont {R.}~\bibnamefont {Reimann}},\ and\ \bibinfo
  {author} {\bibfnamefont {D.}~\bibnamefont {Schneble}},\ }\href
  {https://doi.org/10.1103/PhysRevLett.105.045303} {\bibfield  {journal}
  {\bibinfo  {journal} {Phys. Rev. Lett.}\ }\textbf {\bibinfo {volume} {105}},\
  \bibinfo {pages} {045303} (\bibinfo {year} {2010})}\BibitemShut {NoStop}%
\bibitem [{\citenamefont {Carlini}\ and\ \citenamefont
  {Stringari}(2021)}]{carlini2021}%
  \BibitemOpen
  \bibfield  {author} {\bibinfo {author} {\bibfnamefont {F.}~\bibnamefont
  {Carlini}}\ and\ \bibinfo {author} {\bibfnamefont {S.}~\bibnamefont
  {Stringari}},\ }\href {https://doi.org/10.1103/PhysRevA.104.023301}
  {\bibfield  {journal} {\bibinfo  {journal} {Phys. Rev. A}\ }\textbf {\bibinfo
  {volume} {104}},\ \bibinfo {pages} {023301} (\bibinfo {year}
  {2021})}\BibitemShut {NoStop}%
\bibitem [{\citenamefont {Zheng}\ \emph {et~al.}(2025)\citenamefont {Zheng},
  \citenamefont {Luo}, \citenamefont {Shen}, \citenamefont {He}, \citenamefont
  {Zhu}, \citenamefont {Liu}, \citenamefont {Zhang}, \citenamefont {Sun},
  \citenamefont {Deng}, \citenamefont {Yuan},\ and\ \citenamefont
  {Pan}}]{Zheng2025}%
  \BibitemOpen
  \bibfield  {author} {\bibinfo {author} {\bibfnamefont {Y.-G.}\ \bibnamefont
  {Zheng}}, \bibinfo {author} {\bibfnamefont {A.}~\bibnamefont {Luo}}, \bibinfo
  {author} {\bibfnamefont {Y.-C.}\ \bibnamefont {Shen}}, \bibinfo {author}
  {\bibfnamefont {M.-G.}\ \bibnamefont {He}}, \bibinfo {author} {\bibfnamefont
  {Z.-H.}\ \bibnamefont {Zhu}}, \bibinfo {author} {\bibfnamefont
  {Y.}~\bibnamefont {Liu}}, \bibinfo {author} {\bibfnamefont {W.-Y.}\
  \bibnamefont {Zhang}}, \bibinfo {author} {\bibfnamefont {H.}~\bibnamefont
  {Sun}}, \bibinfo {author} {\bibfnamefont {Y.}~\bibnamefont {Deng}}, \bibinfo
  {author} {\bibfnamefont {Z.-S.}\ \bibnamefont {Yuan}},\ and\ \bibinfo
  {author} {\bibfnamefont {J.-W.}\ \bibnamefont {Pan}},\ }\bibfield  {journal}
  {\bibinfo  {journal} {Nature Physics}\ }\href
  {https://doi.org/10.1038/s41567-024-02732-5} {10.1038/s41567-024-02732-5}
  (\bibinfo {year} {2025})\BibitemShut {NoStop}%
\bibitem [{\citenamefont {Sekino}\ \emph {et~al.}(2023)\citenamefont {Sekino},
  \citenamefont {Tajima},\ and\ \citenamefont {Uchino}}]{sekino2023}%
  \BibitemOpen
  \bibfield  {author} {\bibinfo {author} {\bibfnamefont {Y.}~\bibnamefont
  {Sekino}}, \bibinfo {author} {\bibfnamefont {H.}~\bibnamefont {Tajima}},\
  and\ \bibinfo {author} {\bibfnamefont {S.}~\bibnamefont {Uchino}},\ }\href
  {https://doi.org/10.1103/PhysRevResearch.5.023058} {\bibfield  {journal}
  {\bibinfo  {journal} {Phys. Rev. Res.}\ }\textbf {\bibinfo {volume} {5}},\
  \bibinfo {pages} {023058} (\bibinfo {year} {2023})}\BibitemShut {NoStop}%
\bibitem [{\citenamefont {Pradhan}\ \emph {et~al.}(2024)\citenamefont
  {Pradhan}, \citenamefont {Kanamoto}, \citenamefont {Bhattacharya},\ and\
  \citenamefont {Mishra}}]{pradhan2024}%
  \BibitemOpen
  \bibfield  {author} {\bibinfo {author} {\bibfnamefont {N.}~\bibnamefont
  {Pradhan}}, \bibinfo {author} {\bibfnamefont {R.}~\bibnamefont {Kanamoto}},
  \bibinfo {author} {\bibfnamefont {M.}~\bibnamefont {Bhattacharya}},\ and\
  \bibinfo {author} {\bibfnamefont {P.~K.}\ \bibnamefont {Mishra}},\ }\href
  {https://arxiv.org/abs/2410.21015} {\bibinfo {title} {Signature of
  andreev-bashkin superfluid drag from cavity optomechanics}} (\bibinfo {year}
  {2024}),\ \Eprint {https://arxiv.org/abs/2410.21015} {arXiv:2410.21015
  [cond-mat.quant-gas]} \BibitemShut {NoStop}%
\bibitem [{\citenamefont {Homeier}\ \emph {et~al.}(2024)\citenamefont
  {Homeier}, \citenamefont {Harris}, \citenamefont {Blatz}, \citenamefont
  {Geier}, \citenamefont {Hollerith}, \citenamefont {Schollw\"ock},
  \citenamefont {Grusdt},\ and\ \citenamefont {Bohrdt}}]{homeier2024}%
  \BibitemOpen
  \bibfield  {author} {\bibinfo {author} {\bibfnamefont {L.}~\bibnamefont
  {Homeier}}, \bibinfo {author} {\bibfnamefont {T.~J.}\ \bibnamefont {Harris}},
  \bibinfo {author} {\bibfnamefont {T.}~\bibnamefont {Blatz}}, \bibinfo
  {author} {\bibfnamefont {S.}~\bibnamefont {Geier}}, \bibinfo {author}
  {\bibfnamefont {S.}~\bibnamefont {Hollerith}}, \bibinfo {author}
  {\bibfnamefont {U.}~\bibnamefont {Schollw\"ock}}, \bibinfo {author}
  {\bibfnamefont {F.}~\bibnamefont {Grusdt}},\ and\ \bibinfo {author}
  {\bibfnamefont {A.}~\bibnamefont {Bohrdt}},\ }\href
  {https://doi.org/10.1103/PhysRevLett.132.230401} {\bibfield  {journal}
  {\bibinfo  {journal} {Phys. Rev. Lett.}\ }\textbf {\bibinfo {volume} {132}},\
  \bibinfo {pages} {230401} (\bibinfo {year} {2024})}\BibitemShut {NoStop}%
\bibitem [{\citenamefont {Yanay}\ \emph {et~al.}(2020)\citenamefont {Yanay},
  \citenamefont {Braum\"{u}ller}, \citenamefont {Gustavsson}, \citenamefont
  {Oliver},\ and\ \citenamefont {Tahan}}]{Yanay2020}%
  \BibitemOpen
  \bibfield  {author} {\bibinfo {author} {\bibfnamefont {Y.}~\bibnamefont
  {Yanay}}, \bibinfo {author} {\bibfnamefont {J.}~\bibnamefont
  {Braum\"{u}ller}}, \bibinfo {author} {\bibfnamefont {S.}~\bibnamefont
  {Gustavsson}}, \bibinfo {author} {\bibfnamefont {W.~D.}\ \bibnamefont
  {Oliver}},\ and\ \bibinfo {author} {\bibfnamefont {C.}~\bibnamefont
  {Tahan}},\ }\bibfield  {journal} {\bibinfo  {journal} {npj Quantum
  Information}\ }\textbf {\bibinfo {volume} {6}},\ \href
  {https://doi.org/10.1038/s41534-020-0269-1} {10.1038/s41534-020-0269-1}
  (\bibinfo {year} {2020})\BibitemShut {NoStop}%
\bibitem [{\citenamefont {Harris}\ \emph {et~al.}(2024)\citenamefont {Harris},
  \citenamefont {Schollwöck}, \citenamefont {Bohrdt},\ and\ \citenamefont
  {Grusdt}}]{harris2024kineticmagnetismstripeorder}%
  \BibitemOpen
  \bibfield  {author} {\bibinfo {author} {\bibfnamefont {T.~J.}\ \bibnamefont
  {Harris}}, \bibinfo {author} {\bibfnamefont {U.}~\bibnamefont {Schollwöck}},
  \bibinfo {author} {\bibfnamefont {A.}~\bibnamefont {Bohrdt}},\ and\ \bibinfo
  {author} {\bibfnamefont {F.}~\bibnamefont {Grusdt}},\ }\href
  {https://arxiv.org/abs/2410.00904} {\bibinfo {title} {Kinetic magnetism and
  stripe order in the doped afm bosonic ${t-J}$ model}} (\bibinfo {year}
  {2024}),\ \Eprint {https://arxiv.org/abs/2410.00904} {arXiv:2410.00904
  [cond-mat.quant-gas]} \BibitemShut {NoStop}%
\end{thebibliography}
%

\end{document}